\pdfoutput=1
\documentclass[12pt,a4paper]{article}

\usepackage{ifthen} 
\newboolean{pdflatex}
\setboolean{pdflatex}{true} 

\newboolean{articletitles}
\setboolean{articletitles}{true} 

\newboolean{uprightparticles}
\setboolean{uprightparticles}{false} 

\def\paperauthors{LHCb collaboration} 
\def\paperasciititle{Amplitude analysis and branching fraction measurement of B->DstarDspi decays} 
\def\papertitle{Amplitude analysis and branching fraction measurement of {\boldmath $\Bp\to\Dstarm\Dsp\pip$} decays} 
\def\paperkeywords{{High Energy Physics}, {LHCb}} 
\def\papercopyright{\the\year\ CERN for the benefit of the LHCb collaboration} 
\def\paperlicence{CC BY 4.0 licence}
\def\paperlicenceurl{https://creativecommons.org/licenses/by/4.0/}

\usepackage[top=1in, bottom=1.25in, left=1in, right=1in]{geometry}

\usepackage{microtype}
\usepackage{lineno}  
\usepackage{xspace} 
\usepackage{caption} 

\usepackage{graphicx}  
\usepackage{color}
\usepackage{colortbl}
\graphicspath{{./figs/}} 

\usepackage{amsmath} 
\usepackage{amssymb}
\usepackage{amsfonts}
\usepackage{upgreek} 

\newcommand*\patchAmsMathEnvironmentForLineno[1]{%
\expandafter\let\csname old#1\expandafter\endcsname\csname #1\endcsname
\expandafter\let\csname oldend#1\expandafter\endcsname\csname
end#1\endcsname
 \renewenvironment{#1}%
   {\linenomath\csname old#1\endcsname}%
   {\csname oldend#1\endcsname\endlinenomath}%
}
\newcommand*\patchBothAmsMathEnvironmentsForLineno[1]{%
  \patchAmsMathEnvironmentForLineno{#1}%
  \patchAmsMathEnvironmentForLineno{#1*}%
}
\AtBeginDocument{%
\patchBothAmsMathEnvironmentsForLineno{equation}%
\patchBothAmsMathEnvironmentsForLineno{align}%
\patchBothAmsMathEnvironmentsForLineno{flalign}%
\patchBothAmsMathEnvironmentsForLineno{alignat}%
\patchBothAmsMathEnvironmentsForLineno{gather}%
\patchBothAmsMathEnvironmentsForLineno{multline}%
\patchBothAmsMathEnvironmentsForLineno{eqnarray}%
}

\usepackage{hyperxmp}

\usepackage[pdftex,
            pdfauthor={\paperauthors},
            pdftitle={\paperasciititle},
            pdfkeywords={\paperkeywords},
            pdfcopyright={Copyright (C) \papercopyright},
            pdflicenseurl={\paperlicenceurl}]{hyperref}

\usepackage[colorinlistoftodos,textsize=scriptsize]{todonotes}

\usepackage[bottom,flushmargin,hang,multiple]{footmisc}

\usepackage[all]{hypcap} 

\usepackage{xspace} 
\usepackage{upgreek}

\def\lhcb   {\mbox{LHCb}\xspace}

\def\MagUp {\mbox{\em Mag\kern -0.05em Up}\xspace}

\ifthenelse{\boolean{uprightparticles}}%
{

 \def\Pmu         {\ensuremath{\upmu}\xspace}                 
 \def\Pnu         {\ensuremath{\upnu}\xspace}                 
                  
 \def\Ppi         {\ensuremath{\uppi}\xspace}

 \def\Ptau        {\ensuremath{\uptau}\xspace}

 \def\PDelta      {\ensuremath{\Delta}\xspace}                 
 \def\PXi         {\ensuremath{\Xi}\xspace}                 
 \def\PLambda     {\ensuremath{\Lambda}\xspace}                 
 \def\PSigma      {\ensuremath{\Sigma}\xspace}                 
 \def\POmega      {\ensuremath{\Omega}\xspace}                 
 \def\PUpsilon    {\ensuremath{\Upsilon}\xspace}
 \let\oldPi\Pi
 \def\PPi         {\ensuremath{\oldPi}\xspace}

 \def\PB      {\ensuremath{\mathrm{B}}\xspace}                 
                  
 \def\PD      {\ensuremath{\mathrm{D}}\xspace}

 \def\PK      {\ensuremath{\mathrm{K}}\xspace}

 \def\Pb      {\ensuremath{\mathrm{b}}\xspace}                 
 \def\Pc      {\ensuremath{\mathrm{c}}\xspace}

 \def\Pi      {\ensuremath{\mathrm{i}}\xspace}

 \def\Pp      {\ensuremath{\mathrm{p}}\xspace}

 \def\Ps      {\ensuremath{\mathrm{s}}\xspace}

 \def\thebaroffset{0.0em}
}
{

 \def\Pmu         {\ensuremath{\mu}\xspace}                 
 \def\Pnu         {\ensuremath{\nu}\xspace}                 
                  
 \def\Ppi         {\ensuremath{\pi}\xspace}

 \def\Ptau        {\ensuremath{\tau}\xspace}

 \mathchardef\PDelta="7101
 \mathchardef\PXi="7104
 \mathchardef\PLambda="7103
 \mathchardef\PSigma="7106
 \mathchardef\POmega="710A
 \mathchardef\PUpsilon="7107
 \mathchardef\PPi="7105
                  
 \def\PB      {\ensuremath{B}\xspace}                 
                  
 \def\PD      {\ensuremath{D}\xspace}

 \def\PK      {\ensuremath{K}\xspace}

 \def\Pb      {\ensuremath{b}\xspace}                 
 \def\Pc      {\ensuremath{c}\xspace}

 \def\Pi      {\ensuremath{i}\xspace}

 \def\Pp      {\ensuremath{p}\xspace}

 \def\Ps      {\ensuremath{s}\xspace}

 \def\thebaroffset{0.18em}
}
\newcommand{\offsetoverline}[2][\thebaroffset]{\kern #1\overline{\kern -#1 #2}}%

\makeatletter
\ifcase \@ptsize \relax
  \newcommand{\miniscule}{\@setfontsize\miniscule{4}{5}}
\or
  \newcommand{\miniscule}{\@setfontsize\miniscule{5}{6}}
\or
  \newcommand{\miniscule}{\@setfontsize\miniscule{5}{6}}
\fi
\makeatother

\DeclareRobustCommand{\optbar}[1]{\shortstack{{\miniscule (\rule[.5ex]{1.25em}{.18mm})}
  \\ [-.7ex] $#1$}}

\def\mup        {{\ensuremath{\Pmu^+}}\xspace}

\def\taup       {{\ensuremath{\Ptau^+}}\xspace}

\def\neu        {{\ensuremath{\Pnu}}\xspace}
\def\neub       {{\ensuremath{\overline{\Pnu}}}\xspace}
\def\neum       {{\ensuremath{\neu_\mu}}\xspace}

\def\neut       {{\ensuremath{\neu_\tau}}\xspace}
\def\neutb      {{\ensuremath{\neub_\tau}}\xspace}
\def\neul       {{\ensuremath{\neu_\ell}}\xspace}

\def\squark    {{\ensuremath{\Ps}}\xspace}
\def\squarkbar {{\ensuremath{\overline \squark}}\xspace}

\def\cquark    {{\ensuremath{\Pc}}\xspace}
\def\cquarkbar {{\ensuremath{\overline \cquark}}\xspace}

\def\bquark    {{\ensuremath{\Pb}}\xspace}

\def\pion   {{\ensuremath{\Ppi}}\xspace}
\def\piz    {{\ensuremath{\pion^0}}\xspace}
\def\pip    {{\ensuremath{\pion^+}}\xspace}
\def\pim    {{\ensuremath{\pion^-}}\xspace}

\def\kaon    {{\ensuremath{\PK}}\xspace}

\def\KorKbar {\kern \thebaroffset\optbar{\kern -\thebaroffset \PK}{}\xspace}

\def\Kp      {{\ensuremath{\kaon^+}}\xspace}
\def\Km      {{\ensuremath{\kaon^-}}\xspace}

\def\Dbar    {{\ensuremath{\offsetoverline{\PD}}}\xspace}
\def\D       {{\ensuremath{\PD}}\xspace}
\def\Db      {{\ensuremath{\Dbar}}\xspace}
\def\DorDbar {\kern \thebaroffset\optbar{\kern -\thebaroffset \PD}\xspace}
\def\Dz      {{\ensuremath{\D^0}}\xspace}
\def\Dzb     {{\ensuremath{\Dbar{}^0}}\xspace}
\def\Dp      {{\ensuremath{\D^+}}\xspace}
\def\Dm      {{\ensuremath{\D^-}}\xspace}

\def\DpDm    {\ensuremath{\Dp {\kern -0.16em \Dm}}\xspace}
\def\Dstar   {{\ensuremath{\D^*}}\xspace}

\def\Dstarp  {{\ensuremath{\D^{*+}}}\xspace}
\def\Dstarm  {{\ensuremath{\D^{*-}}}\xspace}

\def\Ds      {{\ensuremath{\D^+_\squark}}\xspace}
\def\Dsp     {{\ensuremath{\D^+_\squark}}\xspace}
\def\Dsm     {{\ensuremath{\D^-_\squark}}\xspace}

\def\Dssp    {{\ensuremath{\D^{*+}_\squark}}\xspace}

\def\B       {{\ensuremath{\PB}}\xspace}

\def\BorBbar {\kern \thebaroffset\optbar{\kern -\thebaroffset \PB}\xspace}
\def\Bz      {{\ensuremath{\B^0}}\xspace}

\def\Bd      {{\ensuremath{\B^0}}\xspace}

\def\BdorBdbar {\kern \thebaroffset\optbar{\kern -\thebaroffset \Bd}\xspace}
\def\Bu      {{\ensuremath{\B^+}}\xspace}
\def\Bub     {{\ensuremath{\B^-}}\xspace}
\def\Bp      {{\ensuremath{\Bu}}\xspace}
\def\Bm      {{\ensuremath{\Bub}}\xspace}

\def\Bs      {{\ensuremath{\B^0_\squark}}\xspace}

\def\BsorBsbar {\kern \thebaroffset\optbar{\kern -\thebaroffset \Bs}\xspace}

\def\Y#1S{\ensuremath{\PUpsilon{(#1S)}}\xspace}

\def\proton      {{\ensuremath{\Pp}}\xspace}

\def\Lz          {{\ensuremath{\PLambda}}\xspace}

\def\LorLbar     {\kern \thebaroffset\optbar{\kern -\thebaroffset \PLambda}\xspace}

\def\Lc          {{\ensuremath{\Lz^+_\cquark}}\xspace}

\def\Lb           {{\ensuremath{\Lz^0_\bquark}}\xspace}

\def\BF         {{\ensuremath{\mathcal{B}}}\xspace}
\def\BR         {\BF}

\newcommand{\decay}[2]{\ensuremath{#1\!\to #2}\xspace} 

\def\to                 {\ensuremath{\rightarrow}\xspace}

\def\CP                {{\ensuremath{C\!P}}\xspace}

\newcommand{\phid}{{\ensuremath{\phi_{\dquark}}}\xspace}

\def\AT#1     {\ensuremath{A_{\mathrm{T}}^{#1}}\xspace}           

\def\C#1      {\ensuremath{\mathcal{C}_{#1}}\xspace}                       
\def\Cp#1     {\ensuremath{\mathcal{C}_{#1}^{'}}\xspace}                    
\def\Ceff#1   {\ensuremath{\mathcal{C}_{#1}^{\mathrm{(eff)}}}\xspace}        
\def\Cpeff#1  {\ensuremath{\mathcal{C}_{#1}^{'\mathrm{(eff)}}}\xspace}       
\def\Ope#1    {\ensuremath{\mathcal{O}_{#1}}\xspace}                       
\def\Opep#1   {\ensuremath{\mathcal{O}_{#1}^{'}}\xspace}                    

\newcommand{\nospaceunit}[1]{\ensuremath{\text{#1}}}       
\newcommand{\aunit}[1]{\ensuremath{\text{\,#1}}}       

\newcommand{\tev}{\aunit{Te\kern -0.1em V}\xspace}
\newcommand{\gev}{\aunit{Ge\kern -0.1em V}\xspace}
\newcommand{\mev}{\aunit{Me\kern -0.1em V}\xspace}
\newcommand{\kev}{\aunit{ke\kern -0.1em V}\xspace}
\newcommand{\ev}{\aunit{e\kern -0.1em V}\xspace}
 
\newcommand{\mevc}{\ensuremath{\aunit{Me\kern -0.1em V\!/}c}\xspace}
\newcommand{\gevc}{\ensuremath{\aunit{Ge\kern -0.1em V\!/}c}\xspace}
\newcommand{\mevcc}{\ensuremath{\aunit{Me\kern -0.1em V\!/}c^2}\xspace}
\newcommand{\gevcc}{\ensuremath{\aunit{Ge\kern -0.1em V\!/}c^2}\xspace}

\def\mum  {\ensuremath{\,\upmu\nospaceunit{m}}\xspace}

\def\fb   {\ensuremath{\aunit{fb}}\xspace}
\def\invfb   {\ensuremath{\fb^{-1}}\xspace}

\newcommand{\chisq}{\ensuremath{\chi^2}\xspace}

\def\gsim{{~\raise.15em\hbox{$>$}\kern-.85em
          \lower.35em\hbox{$\sim$}~}\xspace}
\def\lsim{{~\raise.15em\hbox{$<$}\kern-.85em
          \lower.35em\hbox{$\sim$}~}\xspace}

\def\pt         {\ensuremath{p_{\mathrm{T}}}\xspace}

\def\ptot       {\ensuremath{p}\xspace}

\def\evtgen     {\mbox{\textsc{EvtGen}}\xspace}

\def\geant      {\mbox{\textsc{Geant4}}\xspace}

\def\photos     {\mbox{\textsc{Photos}}\xspace}

\def\pythia     {\mbox{\textsc{Pythia}}\xspace}

\def\tensorflow {\mbox{\textsc{TensorFlow}}\xspace}

\def\tell1  {TELL1\xspace}
\def\ukl1   {UKL1\xspace}

\newcommand{\eg}{\mbox{\itshape e.g.}\xspace}

\newcommand{\lhcborcid}[1]{\href{https://orcid.org/#1}{\hspace*{0.1em}\raisebox{-0.45ex}{\includegraphics[width=1em]{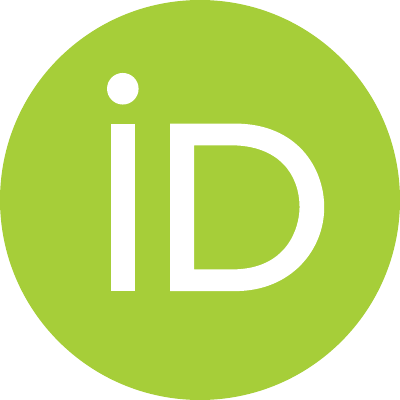}}}}

\usepackage{cite} 
\usepackage{mciteplus}

\usepackage{longtable} 
\usepackage{booktabs}

\usepackage{rotating}

\newcommand{\bdstds}{\ensuremath{\Bz\to\Dstarm\Dsp}\xspace}

\newcommand{\bpdstdspi}{\ensuremath{\Bp\to\Dstarm\Dsp\pip}\xspace}
\newcommand{\bpdstdsstpi}{\ensuremath{\Bp\to\Dstarm\D^{*+}_{\squark}\pip}\xspace}
\newcommand{\dstdspi}{\ensuremath{\Dstarm\Dsp\pip}\xspace}

\newcommand{\msqdstpi}{\ensuremath{m^2(\Dstarm\pip)}\xspace}

\newcommand{\msqdspi}{\ensuremath{m^2(\Dsp\pip)}\xspace}

\newcommand{\costhd}{\ensuremath{\cos\theta_D}\xspace}
\renewcommand{\phid}{\ensuremath{\phi_D}\xspace}
\newcommand{\tcsbar}{\ensuremath{T_{\cquark\squarkbar 0}^{\ast}(2900)^{++}}\xspace}

\def\thetad   {\ensuremath{\theta_{D}}\xspace}

\def\thetadst {\ensuremath{\theta_{D^*}}\xspace}
\newcommand{\hel}[1]{\ensuremath{\lambda_{#1}}}

\newcommand{\wigd}[3]{\ensuremath{d^{#1}_{#2}}\left(#3\right)}

\mathchardef\mhyphen="2D

\begin{document}

\renewcommand{\thefootnote}{\fnsymbol{footnote}}
\setcounter{footnote}{1}


\begin{titlepage}
\pagenumbering{roman}

\vspace*{-1.5cm}
\centerline{\large EUROPEAN ORGANIZATION FOR NUCLEAR RESEARCH (CERN)}
\vspace*{1.5cm}
\noindent
\begin{tabular*}{\linewidth}{lc@{\extracolsep{\fill}}r@{\extracolsep{0pt}}}
\ifthenelse{\boolean{pdflatex}}
{\vspace*{-1.5cm}\mbox{\!\!\!\includegraphics[width=.14\textwidth]{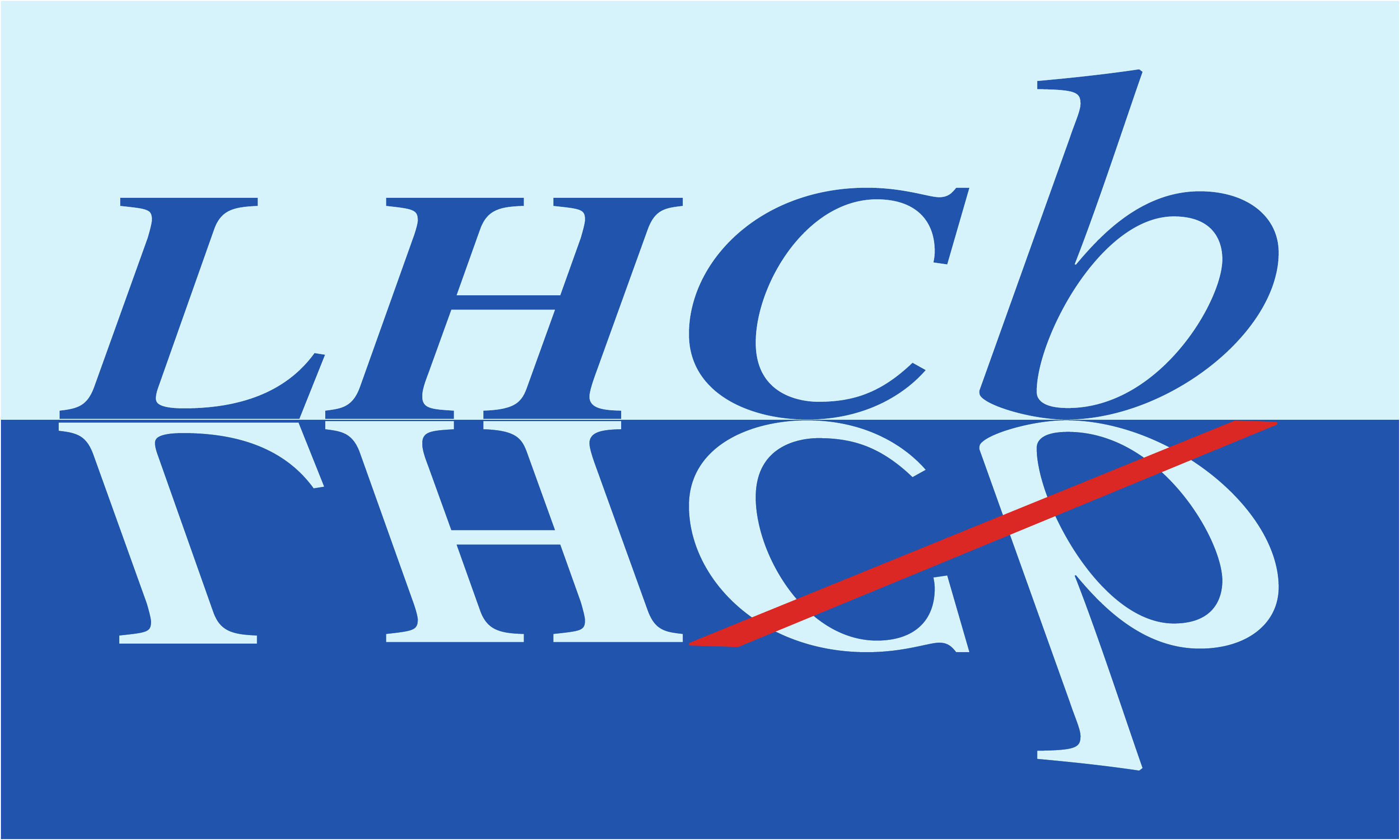}} & &}%
{\vspace*{-1.2cm}\mbox{\!\!\!\includegraphics[width=.12\textwidth]{figs/lhcb-logo.eps}} & &}%
\\
 & & CERN-EP-2024-110 \\  
 & & LHCb-PAPER-2024-001 \\  
 & & September 16, 2024 \\ 
 & & \\
\end{tabular*}

\vspace*{4.0cm}

{\normalfont\bfseries\boldmath\huge
\begin{center}
  \papertitle 
\end{center}
}

\vspace*{2.0cm}

\begin{center}
\paperauthors\footnote{Authors are listed at the end of this paper.}
\end{center}

\vspace{\fill}

\begin{abstract}
  \noindent
  The decays of the $\Bp$ meson to the final state $\Dstarm\Dsp\pip$ are studied in proton-proton collision data collected with the LHCb detector at centre-of-mass energies of 7, 8, and 13\tev, corresponding to a total integrated luminosity of 9\invfb. The ratio of branching fractions of the $\Bp\to\Dstarm\Dsp\pip$ and $\Bz\to\Dstarm\Dsp$ decays is measured to be $0.173\pm 0.006\pm 0.010$, where the first uncertainty is statistical and the second is systematic. Using partially reconstructed $\Dssp\to\Dsp\gamma$ and $\Ds\piz$ decays, the ratio of branching fractions between the $\Bp\to\Dstarm\Dssp\pip$ and $\Bp\to\Dstarm\Dsp\pip$ decays is determined as $1.31\pm 0.07\pm 0.14$. An amplitude analysis of the $\Bp\to\Dstarm\Dsp\pip$ decay is performed for the first time, revealing dominant contributions from known excited charm resonances decaying to the $\Dstarm\pip$ final state. No significant evidence of exotic contributions in the $\Dsp\pip$ or $\Dstarm\Dsp$ channels is found. The fit fraction of the scalar state $\tcsbar$ observed in the $\Bp\to\Dm\Dsp\pip$ decay is determined to be less than 2.3\% at a 90\% confidence level.
\end{abstract}

\vspace*{1.0cm}

\begin{center}
  Published in 
  JHEP 08 (2024) 165
\end{center}

\vspace{\fill}

{\footnotesize 
\centerline{\copyright~\papercopyright. \href{\paperlicenceurl}{\paperlicence}.}}
\vspace*{2mm}

\end{titlepage}


\newpage
\setcounter{page}{2}
\mbox{~}
%
%
%
%


\renewcommand{\thefootnote}{\arabic{footnote}}
\setcounter{footnote}{0}

\cleardoublepage


\pagestyle{plain} 
\setcounter{page}{1}
\pagenumbering{arabic}


\section{Introduction}
\label{sec:Introduction}

Recent studies of \B-meson decays involving pairs of open-charm hadrons have yielded a number of intriguing results in the field of open-charm and charmonium spectroscopy. The LHCb collaboration has conducted a number of these studies, including the analysis of $\Bp\to\Dp\Dm\Kp$ decays\footnote{Charge conjugation is implied throughout this paper unless explicitly stated.} that led to the discovery of tetraquark states in the $\Dm\Kp$ system~\cite{LHCB-PAPER-2020-024, LHCB-PAPER-2020-025}, the observation of near-threshold $\Dsp\Dsm$ structures in the analysis of $\Bp\to\Dsp\Dsm\Kp$ decays~\cite{LHCB-PAPER-2022-018, LHCB-PAPER-2022-019}, and the first observation of a doubly charged charm tetraquark and its neutral partner in $\Bp\to\Dm\Dsp\pip$ and $\Bz\to\Dzb\Dsp\pim$ decays~\cite{LHCb-PAPER-2022-026, LHCb-PAPER-2022-027}. 

In addition to the primary objective of studying doubly charmed decays to improve the understanding of the strong interaction, the analyses of doubly charmed final states can provide valuable information for the studies of semileptonic decays used to search for phenomena beyond the Standard Model. Semileptonic decays involving higher excitations of \D mesons decaying to $\Dstarm\pi$ final state, dominated by the first orbital excitations denoted as $\D^{**}$, constitute a substantial background to $\Bz\to\Dstarm\ell^+\neul$ decays. Thus, the knowledge of the spectrum of these excitations is essential for the measurements of such quantities as the lepton flavour universality ratio $R(\Dstar)$ (see, \eg \cite{LHCb-PAPER-2022-039, LHCb-PAPER-2022-052}). The properties of excited charm mesons decaying to $\Dstarp\pim$ final state have been recently studied in the analysis of $\Bm\to\Dstarp\pim\pim$ decays~\cite{LHCB-PAPER-2019-027}. However, a reliable prediction of the $\B\to \D^{**}$ form factor in semileptonic decays requires additional information from other hadronic decays, such as $\B\to \D^{**}\Dsp$ and $\B\to\D^{**}\D_s^{*+}$ processes~\cite{LeYaouanc:2022dmc}. 

Decays of \B mesons to double-charm final states, themselves, constitute an essential background to the semileptonic \B decays. For instance, processes in which one of the charm mesons decays to a final state with a muon are a background to $\Bz\to\Dstarm\mup\neum$ decays, and to $\Bz\to\Dstarm\taup\neut$ decays where the $\taup$ lepton is reconstructed in the $\taup\to\mup\neum\neutb$ final state. Double-charm decays with one of the charm mesons (typically, the $\Dsp$ meson) decaying to $\pip\pim\pip$ are a background to the $\Bz\to\Dstarm\taup\neut$ processes with the subsequent $\taup\to\pip\pim\pip(\piz)\neutb$ transition. 

The dominant weak transition leading to the double-charm final states is $\bquark\to\cquark\cquarkbar\squark$, with either internal $W$ emission topology and intermediate $\cquark\cquarkbar$ resonances decaying further to $\D\Dbar$ state, or external $W$ emission with intermediate $\Ds$ or $\D$ resonances. While the $\D\Db K$ final states have been studied by both $B$-factories and LHCb~\cite{PhysRevD.68.092001,PhysRevD.83.032004,PhysRevLett.100.092001,PhysRevD.91.052002,LHCb-PAPER-2020-006}, experimental information on the $\B\to \D^{(*)}\D_s^{(*)+}$ decays is still scarce. Notably, evidence of the $\Bp\to\Db^{**0}\D_{s}^{(*)+}$ decay was reported by the CLEO collaboration~\cite{CLEO:2000svj}. In addition to the amplitude analyses of the $\Bp\to\Dm\Dsp\pip$ and the $\Bz\to\Dzb\Dsp\pim$ decays~\cite{LHCb-PAPER-2022-026, LHCb-PAPER-2022-027} mentioned above, an angular analysis of the $\Bz\to\Dstarm\D_s^{*+}$ decay has been performed by LHCb~\cite{LHCB-PAPER-2021-006}.

This paper presents the observation and measurement of the branching fraction of the decay $\bpdstdspi$, along with an analysis of its amplitude structure. A measurement of the branching fraction of the $\Bp\to\Dstarm\Dssp\pip$ decay is also reported using the $\Bz\to\Dstarm\Dsp$ decay as a normalisation channel. The analysis is based on the proton-proton ($\proton\proton$) collision data collected with the LHCb detector at centre-of-mass energies of 7 and 8\tev (Run 1), and 13\tev (Run 2), corresponding to a total luminosity of approximately 9\invfb.

\section{Amplitude analysis formalism}
\label{sec:Formalism}

The process \bpdstdspi is a three-body decay of a pseudoscalar particle to two pseudoscalar and one vector final-state particles. The differential decay rate can be written as
\begin{equation}
  {\rm d}\Gamma = \left|\mathcal{A}(\msqdstpi, \msqdspi, \theta_{\D}, \phid)\right|^{2}\; {\rm d}\msqdstpi\; {\rm d}\msqdspi\; {\rm d}\costhd\; {\rm d}\phid, 
  \label{eq:decayrate}
\end{equation}
where $\mathcal{A}$ is the decay amplitude, \msqdstpi and \msqdspi are the squared invariant masses of the $\Dstarm\pip$ and $\Dsp\pip$ combinations, respectively, and $\theta_D$ and $\phid$ are the polar and azimuthal angles of the \Dzb meson in the rest frame of the \Dstarm particle.

\subsection{Amplitude model}
\label{subsec:AmpModel}
Within the isobar model used in this analysis, the decay amplitude $\mathcal{A}$ is expressed as a sum of quasi-two-body amplitudes with potential resonant and nonresonant intermediate states in the $\Dstarm\pip$, $\Dsp\pip$ and $\Dstarm\Dsp$ channels: 
\begin{equation}
  \mathcal{A} = \mathcal{A}^{(\Dstar\pi)} + \mathcal{A}^{(\D_s\pi)} + \mathcal{A}^{(\Dstar\D_s)}. 
  \label{eq:amplitude}
\end{equation}
Each of the amplitudes is represented as a sum over the resonant or nonresonant components, which in turn are the products of the line shape $\mathcal{R}_n$ and the angular $\mathcal{H}_n$ terms, where $n$ denotes the index of the component in each quasi-two-body decay channel
(${\rm ch} = \Dstarm\pip, \Dsp\pip, \Dstarm\Dsp$): 
\begin{equation}
    \mathcal{A}^{\rm (ch)} = \sum_{n}\mathcal{A}_n^{\rm (ch)} = 
    \sum_{n}\mathcal{R}^{\rm (ch)}_{n}\mathcal{H}^{\rm (ch)}_{n}. 
\end{equation}

\begin{figure}[t]
  \begin{center}
  \includegraphics[width=0.32\textwidth]{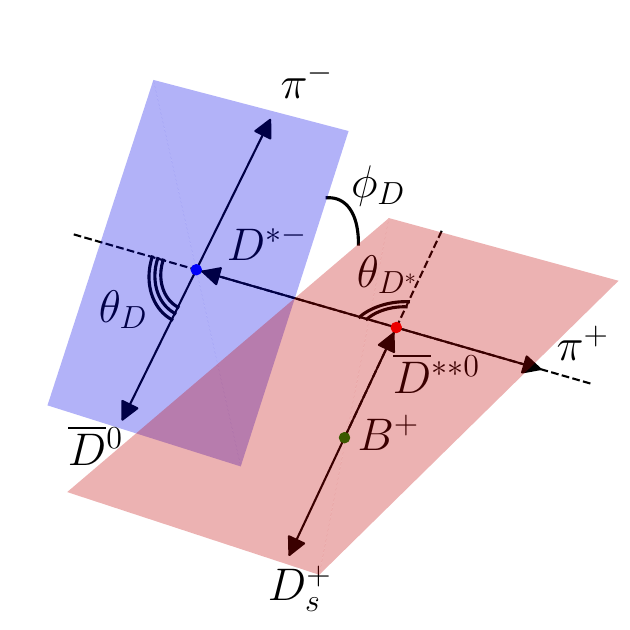}
  \put(-135, 110){(a)}
  \raisebox{-0.02\textwidth}{
    \includegraphics[width=0.32\textwidth]{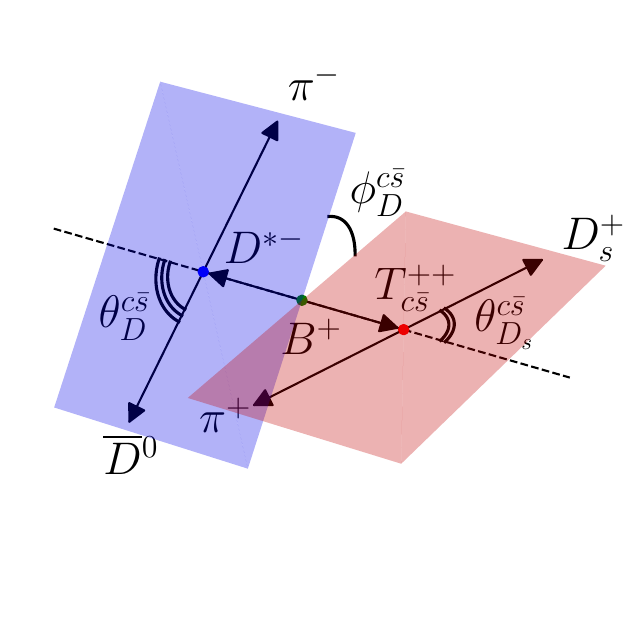}
  }
  \put(-135, 110){(b)}
  \includegraphics[width=0.32\textwidth]{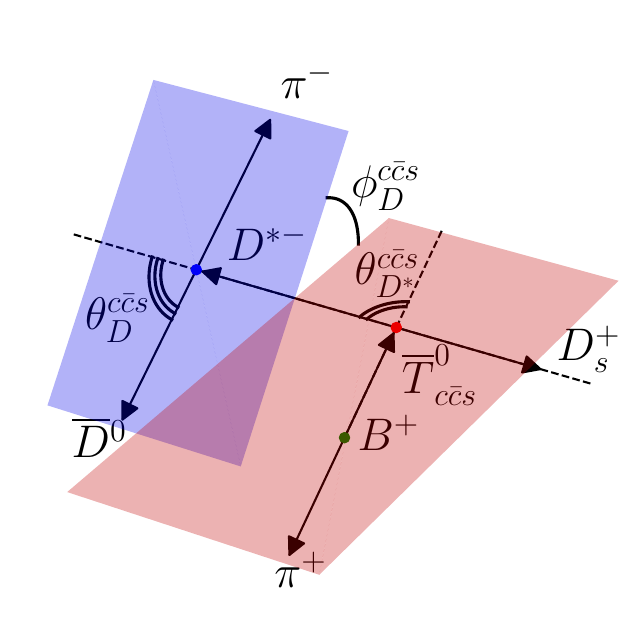}
  \put(-135, 110){(c)}
  \end{center}
  \vspace{-0.06\textwidth}
  \caption{Definition of the helicity angles in the (a) $\Bp\to \Db^{**0}(\to\Dstarm\pip)\Dsp$, 
           (b) \mbox{$\Bp\to \Dstarm T_{c\bar{s}}^{++}(\to\Dsp\pip)$}, and 
           (c) $\Bp\to \overline{T}_{c\bar{c} s}^0(\to\Dstarm\Dsp)\pip$ decay chains. Here $T_{c\bar{s}}^{++}$ and $\overline{T}_{c\bar{c} s}^0$ are the hypothetical tetraquark-like states decaying to $\Dsp\pip$ and $\Dstarm\Dsp$ final states, respectively, following the nomenclature suggested in Ref.~\cite{PDGHadrons}. }
  \label{fig:angles}
\end{figure}

The angular terms are functions of the helicity angles, defined for each channel in Fig.~\ref{fig:angles} (see, \eg, Ref.~\cite{Chung:1971ri} for the formalism used): 
\begin{equation}
  \begin{split}
    \mathcal{H}^{(\Dstar\pi)}_n & = \sum_{\lambda=0,\pm 1} h^{(\Dstar\pi)}_{n,\lambda}\;
                    \wigd{J_n}{0, \lambda}{\thetadst}\; 
                    \wigd{1}{\lambda, 0}{\thetad}\;
                    \exp{\left(i\lambda\phid\right)}, \\                
    \mathcal{H}^{(\D_s\pi)}_n  & = \sum_{\lambda=0,\pm 1} h^{(\D_s\pi)}_{n,\lambda}\;
      \wigd{J_{n}}{\lambda, 0}{\theta_{\D_s}^{c\bar{s}}} \; 
      \wigd{1}{\lambda, 0}{\thetad^{c\bar{s}}} \; 
      \exp{\left(i\lambda\phid^{c\bar{s}}\right)},  \\
    \mathcal{H}^{(\Dstar\D_s)}_n & = \sum_{\lambda=0,\pm 1} h^{(\Dstar\D_s)}_{n,\lambda}\;
      \wigd{J_{n}}{0, \lambda}{\thetadst^{c \bar{c} s}} \; 
      \wigd{1}{\lambda, 0}{\thetad^{c\bar{c} s}} \; 
      \exp{\left(i\lambda\phid^{c \bar{c} s}\right)}.
  \end{split}
  \label{eq:helamp_dstpi}
\end{equation}
Here $\lambda$ is the helicity of the $\Dstarm$ meson, $J_n$ is the spin of the intermediate resonance, $d^{J}_{\lambda,\lambda^\prime}(\theta)$ are the reduced Wigner functions, and $h_{n,\lambda}$ are complex couplings for each helicity amplitude. The angles $\theta_D$ and $\phi_D$ in these expressions are the same as those used as the phase-space variables in Eq.~\ref{eq:decayrate}, while the other angles are functions of the four phase-space variables. The angles $\theta_D^{(c\bar{s},c\bar{c} s)}$ and $\phi_{D}^{(c\bar{s},c\bar{c} s)}$ are defined in the $\Dstarm$ rest frame, while $\thetadst$, $\thetadst^{c\bar{c} s}$ and $\theta_{\D_s}^{c\bar{s}}$ are defined in the rest frames of the $\Db^{**0}$, $\overline{T}_{c\bar{c} s}^0$, and $T_{c\bar{s}}^{++}$ states, respectively. 

The amplitudes of the two-body decay $A\to BC$ can be expressed in terms of $LS$ couplings $c_{L,S}$, where $L$ and $S$ denote the orbital angular momentum and the total spin of the $BC$ combination. The relations between the helicity couplings and the $LS$ couplings are given by Wigner 3-$\mathrm{j}$ symbols, 
\begin{gather}
	h^{(\decay{A}{BC})}_{\hel{B},\hel{C}} = c_{L,S}\sqrt{\frac{2L+1}{2J_{A}+1}}
	\left(
	\begin{matrix}
		J_{B}   & J_{C}   \\
		\hel{B} &-\hel{C} \\
	\end{matrix}
	\;\;
	\begin{matrix}
		S \\
		\hel{B}-\hel{C}
	\end{matrix}
	\right)
	\left(
	\begin{matrix}
		L & S\\
		0 & \hel{B}-\hel{C}\\
	\end{matrix}
	\;\;
	\begin{matrix}
		J_{A} \\
		\hel{B}-\hel{C}
	\end{matrix}
	\right),
\end{gather}
with $J_{A,B,C}$ being the spins, and $\lambda_{A,B,C}$ the helicities of the corresponding particles.

In the case of the $\Bp\to\Db^{**0}(\to\Dstarm\pip)\Dsp$ decay chain,\footnote{Although usually $\D^{**}$ denotes only the first orbital excitations of charm mesons, here this notation is used for any intermediate $\Dstarm\pip$ state.} the helicity of the $\Db^{**0}$ state is always zero and the angular momentum $L_B$ of the $\Bp\to \Db^{**0}\Dsp$ decay is equal to the spin $J$ of the $\Db^{**0}$ resonance. The total spin $S$ of the $\Dstarm\pip$ combination is 1, and the possible values of the orbital momentum $L_R$ for the $\Db^{**0}\to \Dstarm\pip$ decay are constrained by the spin $J$ and parity $P$ of the intermediate $\Db^{**0}$ state. The expression for the coupling with the $\Dstarm$ helicity $\lambda$ reduces to 
\begin{gather}\label{eq:LScoup_dstpi}
	h_{n,\lambda}^{(R\to \Dstar\pi)} = c_{L_R,1}^{(n)}\sqrt{\frac{2L_R+1}{2J+1}}
	\left(
	\begin{matrix}
		1   & 0 & 1\\
		\lambda &0 & \lambda\\
	\end{matrix}
	\right)
	\left(
	\begin{matrix}
		L_R & 1 & J \\
		0 & \lambda & \lambda\\
	\end{matrix}
	\right). 
\end{gather}
Combining this with Eq.~\ref{eq:helamp_dstpi}, the angular terms of the amplitude given in Table~\ref{tab:hel_amplitudes_dstpi} are obtained. The expressions for the $h_{n,\lambda}^{(R\to \Dstar\D_s)}$ couplings in the $\Bp\to \overline{T}_{c\bar{c} s}^0(\to\Dstarm\Dsp)\pip$ decay chain are identical, with the substitution of the angles $\theta_{\Dstar}$, $\theta_D$, and $\phi_D$ by $\theta_{\Dstar}^{c\bar{c} s}$, $\theta_D^{c\bar{c} s}$, and $\phi_D^{c\bar{c} s}$, respectively. 

In the decay chain $\Bp\to\Dstarm T_{c\bar{s}}^{++}(\to \Dsp\pip)$, the helicities of the $\Dstarm$ and $T_{c\bar{s}}^{++}$ states are equal, their total spin is zero, and the range of angular momenta $L_B$ depends on the spin $J$ of the $T_{c\bar{s}}^{++}$ state: $|J-1|\leq L_B\leq J+1$. The expression for the helicity coupling takes the form
\begin{gather}\label{eq:LScoup_dspi}
	h_{n,\lambda}^{(\B\to \Dstar R)} = c_{L_B,L_B}^{(n)}\sqrt{2L_B + 1}
	\left(
	\begin{matrix}
		J   & 1 & L_B \\
		\lambda & -\lambda & 0\\
	\end{matrix}
	\right)
	\left(
	\begin{matrix}
		L_B & L_B & 0 \\
		0 & 0 & 0\\
	\end{matrix}
	\right),
\end{gather}
while in the $T_{c\bar{s}}^{++}\to \Dsp\pip$ decay the angular momentum is fixed by the spin $J$, and parity conservation requires that the parity $P$ of the $T_{c\bar{s}}^{++}$ state equals $(-1)^J$. The explicit expressions for the angular dependency in this case are given in Table~\ref{tab:hel_amplitudes_dspi}. 

\begin{table}
  \caption{\small The angular dependencies for the $\Bp\to\Db^{**0}(\to \Dstarm\pip)\Dsp$ partial-wave terms $\mathcal{H}_n^{(\Dstar\pi)}$. }
  \label{tab:hel_amplitudes_dstpi}
  \begin{center}
  \begin{tabular}{c|c|c}
        $J^P$ & $L_R$ & Angular term \\
        \hline
        $0^-$ & $1$ & $-\cos\theta_D$ \\
        $1^-$ & $1$ & $\frac{i}{\sqrt{2}}\sin\theta_{D^*}\sin\theta_D\sin\phi_D$ \\
        $1^+$ & $0$ & $\frac{1}{\sqrt{3}}(\cos\theta_{D^*}\cos\theta_D-\sin\theta_{D^*}\sin\theta_D\cos\phi_D)$ \\
              & $2$ & $-\frac{1}{\sqrt{6}}(2\cos\theta_{D^*}\cos\theta_D+\sin\theta_{D^*}\sin\theta_D\cos\phi_D)$ \\
        $2^-$ & $1$ & $\frac{1}{\sqrt{10}}(-3\sin\theta_{D^*}\cos\theta_{D^*}\sin\theta_D\cos\phi_D-3\sin^2\theta_{D^*}\cos\theta_D+2\cos\theta_D)$ \\
              & $3$ & $\frac{\sqrt{15}}{10}(-2\sin\theta_{D^*}\cos\theta_{D^*}\sin\theta_D\cos\phi_D+3\sin^2\theta_{D^*}\cos\theta_D-2\cos\theta_D)$ \\
        $2^+$ & $2$ & $\frac{i\sqrt{6}}{2}\sin\theta_{D^*}\cos\theta_{D^*}\sin\theta_D\sin\phi_D$ \\
  \end{tabular}
  \end{center}
\end{table}

\begin{table}
  \caption{\small The angular dependencies for the $\Bp\to\Dstarm T_{c\bar{s}}^{++}(\to \Dsp\pip)$ partial-wave terms $\mathcal{H}_n^{(\D_s\pi)}$. }
  \label{tab:hel_amplitudes_dspi}
  \begin{center}
  \begin{tabular}{c|c|c}
        $J^P$ & $L_B$ & Angular term \\
        \hline
        $0^+$ & $1$ & $-\cos\theta^{c\bar{s}}_D$ \\
        $1^-$ & $0$ & $-\frac{1}{\sqrt{3}}(\cos\theta^{c\bar{s}}_{\D_s}\cos\theta^{c\bar{s}}_D-\sin\theta^{c\bar{s}}_{\D_s}\sin\theta^{c\bar{s}}_D\cos\phi^{c\bar{s}}_D)$ \\
              & $1$ & $-\frac{i}{\sqrt{2}}\sin\theta^{c\bar{s}}_{\D_s}\sin\theta^{c\bar{s}}_D\sin\phi^{c\bar{s}}_D$ \\
              & $2$ & $\frac{1}{\sqrt{6}}(2\cos\theta^{c\bar{s}}_{\D_s}\cos\theta^{c\bar{s}}_D+\sin\theta^{c\bar{s}}_{\D_s}\sin\theta^{c\bar{s}}_D\cos\phi^{c\bar{s}}_D)$ \\
        $2^+$ & $1$ & $\frac{1}{\sqrt{10}}(-3\sin\theta^{c\bar{s}}_{\D_s}\cos\theta^{c\bar{s}}_{\D_s}\sin\theta^{c\bar{s}}_D\cos\phi^{c\bar{s}}_D-3\sin^2\theta^{c\bar{s}}_{\D_s}\cos\theta^{c\bar{s}}_D+2\cos\theta^{c\bar{s}}_D)$ \\
              & $2$ & $\frac{i\sqrt{6}}{2}\sin\theta^{c\bar{s}}_{\D_s}\cos\theta^{c\bar{s}}_{\D_s}\sin\theta^{c\bar{s}}_D\sin\phi^{c\bar{s}}_D$ \\
              & $3$ & $\frac{\sqrt{15}}{10}(-2\sin\theta^{c\bar{s}}_{\D_s}\cos\theta^{c\bar{s}}_{\D_s}\sin\theta^{c\bar{s}}_D\cos\phi^{c\bar{s}}_D+3\sin^2\theta^{c\bar{s}}_{\D_s}\cos\theta^{c\bar{s}}_D-2\cos\theta^{c\bar{s}}_D)$ \\
  \end{tabular}
  \end{center}
\end{table}

The resonant states are parametrised by relativistic Breit--Wigner (BW) line shapes of the form
\begin{equation}
	\mathcal{R}_{\rm BW}(m) = \left(\frac{q(m)}{q_{0}}\right)^{L_R}\left(\frac{p(m)}{p_{0}}\right)^{L_B}\frac{F_R(m,L_R)F_B(m,L_B)}{m_{0}^{2}-m^{2}-i m_{0}\Gamma(m)}, 
\end{equation}
with the mass-dependent width, $\Gamma(m)$, given by
\begin{equation}
	\Gamma(m) = \Gamma_{0}\left(\frac{q(m)}{q_{0}}\right)^{2L_R+1}\frac{m_{0}}{m}F_{R}^{2}(m,L_R),
\end{equation}
where $m$ is the invariant mass of the resonance decay products, $m_{0}$ and $\Gamma_{0}$ are, respectively, the mass and the width of the intermediate resonance, $L_B$ is the orbital momentum of the $B$ decay, $L_R$ is the orbital momentum of the decay of the resonance, while $p$ and $q$ are, respectively, the momentum of the resonance in the $\Bp$ rest frame and the momentum of resonance decay products in its rest frame. The Blatt--Weisskopf form factors~\cite{Blatt:1952ije} for the intermediate resonance, $F_{R}(m,L_R)$, and for the \Bp meson, $F_{\B}(m,L_B)$, are parametrised as
\begin{equation}
  F_{R,B}\left(m,L\right) = \begin{cases}
		1 & \mbox{for } L = 0, \\
		\sqrt{\frac{1+z^{2}(m)}{1+z_{0}^{2}}} & \mbox{for } L = 1,\\
		\sqrt{\frac{9+3z^{2}(m)+z^{4}(m)}{9+3z^{2}_{0}+z^{4}_{0}}} & \mbox{for } L = 2,\\
		\sqrt{\frac{225+45z^{2}(m)+6z^{4}(m)+z^{6}}{225+45z^{2}_{0}+6z^{4}_{0}+z_{0}^{6}}} & \mbox{for } L = 3,
  \end{cases}
\end{equation}
where $z(m) = p(m)d$, $z_{0}=p_{0}d$, $q_0$ and $p_0$ are the momenta evaluated at the nominal resonance mass, and $d$ is the radial parameter set to $4.5\gev^{-1}$~\cite{LHCB-PAPER-2019-027} for all resonances.\footnote{Natural units with $c=\hbar=1$ are used throughout this paper.}

Nonresonant contributions are parametrised using an exponential function,
\begin{equation}
    \mathcal{R}_{\rm NR}\left(m\right) = \left(\frac{q(m)}{q_{0}}\right)^{L_{R}}\left(\frac{p(m)}{p_{0}}\right)^{L_{\B}}e^{-\alpha(m^2-m_0^2)},
\end{equation}
where $\alpha$ is a shape parameter that is extracted from the fit. The value of $m_0$ in this expression is arbitrary and only affects the numerical values of the couplings. When adding nonresonant \Dstarm\pip or \Dsp\pip amplitudes, $m_0$ is set to $2.15\gev$, while for nonresonant \Dstarm\Dsp amplitudes, $m_{0} = 4\gev$, which roughly corresponds to kinematic thresholds of the respective channels. 

\subsection{Amplitude fit procedure}
\label{subsec:AmpFit}
The probability density function (PDF) used in the amplitude analysis consists of signal and background contributions and is a function of the phase-space variables $\mathbf{x} \equiv \{m^2(\Dstarm\pip), m^2(\Dsp\pip), \theta_D, \phi_D\}$. The amplitude fit minimises the unbinned negative logarithmic likelihood (NLL)
\begin{equation}
    -\ln\mathcal{L} = - \sum_{i=1}^{N}\ln\mathcal{P}_{\rm tot}(\mathbf{x}_i),
    \label{eq:nll}
\end{equation}
where $N$ is the total number of candidates and $\mathcal{P}_{\rm tot}(\mathbf{x}_i)$ is the total PDF for candidate $i$, including the efficiency term and the background component
\begin{equation}
    \mathcal{P}_{\rm tot}(\mathbf{x}) = \left\vert\mathcal{A}(\mathbf{x})\right\vert^{2}\epsilon(\mathbf{x})\frac{1-f_{\rm bkg}}{\mathcal{N}_{\rm sig}}+\mathcal{P}_{\rm bkg}(\mathbf{x})\frac{f_{\rm bkg}}{\mathcal{N}_{\rm bkg}}.
    \label{eq:totalpdf}
\end{equation}
Explicit parametrisations are used to describe the efficiency $\epsilon$, and the background density $\mathcal{P}_{\rm bkg}$ as functions of $\mathbf{x}$ (see Sects.~\ref{sec:Efficiency} and \ref{sec:Backgrounds}). The parameter $f_{\rm bkg}$ is the fraction of background events in the \Bp signal region and is fixed to the value obtained from the results of the fits to the invariant-mass distributions described in Sect.~\ref{sec:Massfit}. The signal $\mathcal{N}_{\rm sig}$ and background $\mathcal{N}_{\rm bkg}$ normalisation integrals are calculated numerically using a sample of $10^6$ events distributed uniformly over the four-dimensional phase space of the decay. At each minimisation step, the PDF is evaluated on the normalisation sample, and the integrals are calculated by summing the density for each normalisation event. The signal decay amplitude $\mathcal{A}$ in Eq.~\ref{eq:amplitude} is constructed using the \textsc{AmpliTF} package~\cite{ampliTF}. The amplitude fit is implemented in \textsc{TFA2}~\cite{tfa2}, a fitting package based on \tensorflow~\cite{tensorflow2015-whitepaper}, interfaced with the \textsc{iminuit} minimisation library~\cite{iminuit}. 

The fit fraction $\mathcal{F}_{i}$ due to a resonant or nonresonant contribution $i$ is computed from the fitted parameters, and is defined as 
\begin{equation}\label{eq:fitfrac}
	\mathcal{F}_{i} = \frac{\int \vert \mathcal{A}_{i}\left(\mathbf{x}\right) \vert^{2}\; {\rm d}\mathbf{x}}{\int \vert \sum_{j} \mathcal{A}_{j} \left(\mathbf{x}\right) \vert^{2}\; {\rm d}\mathbf{x}}.
\end{equation}
If all components correspond to partial waves with different quantum numbers, the sum of the fit fractions adds up to 100\%; otherwise, the sum may differ from 100\% due to interference effects. 

\section{Detector and simulation}
\label{sec:Detector}

The \lhcb detector~\cite{LHCb-DP-2008-001,LHCb-DP-2014-002} is a single-arm forward
spectrometer covering the \mbox{pseudorapidity} range $2<\eta <5$,
designed for the study of particles containing \bquark or \cquark
quarks. The detector includes a high-precision tracking system
consisting of a silicon-strip vertex detector surrounding the $pp$
interaction region, a large-area silicon-strip detector located
upstream of a dipole magnet with a bending power of about
$4{\mathrm{\,Tm}}$, and three stations of silicon-strip detectors and straw drift tubes placed downstream of the magnet.
The tracking system provides a measurement of the momentum, \ptot, of charged particles with a relative uncertainty that varies from 0.5\% at low momentum to 1.0\% at 200\gev. 
The impact parameter (IP), defined as the minimum distance of a track to a primary $pp$ collision vertex (PV), is measured with a resolution of $(15+29/\pt)\mum$, where \pt is the component of the momentum transverse to the beam, in\,\gev.
Different types of charged hadrons are distinguished using information
from two ring-imaging Cherenkov (RICH) detectors. 
Photons, electrons and hadrons are identified by a calorimeter system consisting of scintillating-pad and preshower detectors, an electromagnetic and a hadronic calorimeter. Muons are identified by a
system composed of alternating layers of iron and multiwire
proportional chambers.

The online event selection is performed by a trigger~\cite{LHCb-DP-2012-004,LHCb-DP-2019-001}, which consists of a hardware stage, based on information from the calorimeter and muon systems, followed by a software stage, which applies a full event reconstruction. At the hardware trigger stage, events are required to have a muon with high \pt or a hadron, photon or electron with high transverse energy in the calorimeters. The software trigger requires a two-, three- or four-track secondary vertex with a significant displacement from any primary $pp$ interaction vertex. At least one charged particle must have a high transverse momentum $\pt$ and be inconsistent with originating from a PV. A multivariate algorithm~\cite{BBDT,LHCb-PROC-2015-018} is used for the identification of secondary vertices consistent with the decay of a \bquark hadron.

Simulation is required to model the effects of the detector acceptance and the imposed selection requirements. In the simulation, $pp$ collisions are generated using \pythia~\cite{Sjostrand:2007gs,*Sjostrand:2006za} with a specific \lhcb configuration~\cite{LHCb-PROC-2010-056}. Decays of unstable particles are described by \evtgen~\cite{Lange:2001uf}, in which final-state radiation is generated using \photos~\cite{davidson2015photos}. The interaction of the generated particles with the detector, and its response, are implemented using the \geant toolkit~\cite{Allison:2006ve, Agostinelli:2002hh} as described in Ref.~\cite{LHCb-PROC-2011-006}. 

A data-driven approach is employed to correct the simulated particle identification (PID) information based on large samples of $\Dstarp\to\Dz\pip$, $\Dz\to\Km\pip$ events. For each track PID response, unbinned four-dimensional probability density functions are extracted for data, $p_{\rm data}(x\vert\;\pt, \eta, N_{\rm tr})$, and simulated samples, $p_{\rm sim}(x\vert\;\pt, \eta, N_{\rm tr})$ based on a kernel density estimation technique~\cite{Poluektov:2014rxa}, where $x$ is the PID response, $\pt$ and $\eta$ are the transverse momentum and pseudorapidity of the track, and $N_{\rm tr}$ is the track multiplicity of the event. The cumulative distribution functions for data, $P_{\rm data}(x\vert\;\pt, \eta, N_{\rm tr})$ and for the simulated samples $P_{\rm sim}(x\vert\;\pt, \eta, N_{\rm tr})$ are determined, and the corrected PID response $x_{\rm corr}$ is evaluated by transforming the simulated response $x_{\rm sim}$ as \mbox{$x_{\rm corr} = P^{-1}_{\rm data}(P_{\rm sim}(x_{\rm sim}\vert\;\pt, \eta, N_{\rm tr})\vert\;\pt, \eta, N_{\rm tr})$}~\cite{Aaij:2018vrk}.  

\section{Signal selection}
\label{sec:Selection}

Two decay chains are reconstructed: the \bpdstdspi decays are selected for the amplitude analysis and branching fraction measurement, while the \bdstds decays are used as the normalisation sample for the \bpdstdspi branching fraction measurement. The $\Dzb$ candidates are reconstructed in the $\Kp\pim$ final state, \Dstarm candidates are formed from the $\Dzb\pim$ combinations, and \Dsp candidates are reconstructed in the $\Kp\Km\pip$ final state. Pion and kaon candidate tracks used to create combinations are selected based on loose requirements for track-fit quality, momentum $p$, and transverse momentum \pt. They are required to lie within the RICH detector acceptance and be positively identified as either a pion or a kaon, respectively. To suppress the combinatorial background from the tracks originating from the PV, all tracks are required to be displaced from any PV in the event. The separation from the PV is characterised by the quantity $\chisq_{\rm IP}$, defined as the difference in the vertex-fit \chisq of a given PV reconstructed with and without the track under consideration.

The \Dsp and \Dzb candidates are required to form good-quality vertices well separated from any PV. Their invariant masses must be within 20\mev from the known masses of the respective particles~\cite{PDG2022}. The \Dstarm candidates are formed from the combinations of the \Dzb candidate and a pion track creating a good-quality vertex and having the difference of the invariant masses of \Dstarm and \Dzb candidates less than 150\mev. When constructing the \bpdstdspi and \bdstds decay candidates, a kinematic fit is performed~\cite{Hulsbergen:2005pu}, which constrains the \B and $\D_{(\squark)}$ meson decay products to originate from the corresponding vertices and the masses of the \Dzb and \Dsp candidates to match their known values~\cite{PDG2022}. The vertices of the \Dzb and \Dsp candidates are required to be downstream of the fitted \B meson vertex. In the calculation of the phase-space variables $\msqdstpi$, $\msqdspi$, $\costhd$ and $\phid$, the mass of the $\Dstarm\Dsp\pip$ combination is further constrained in the kinematic fit to be equal to the known \Bp meson mass $m_{\Bp}$~\cite{PDG2022}. 

In the selection, trigger signals are associated with reconstructed particles. Selection requirements are made on whether the trigger decision was due to the signal candidate (trigger-on-signal, or TOS, category), other particles produced in the $pp$ collision (trigger independent of signal, or TIS, category), or both. The candidates in the analysis are categorised as TOS and NotTOS (which comprises the candidates selected as TIS, but not as TOS) based on the hardware trigger decision. Subsequent selection requirements, including software triggers, do not induce further splits in the analysed data sample.  

A boosted decision tree (BDT) classifier~\cite{Breiman,AdaBoost} is used to separate signal candidates from the background arising from random combinations of final-state particles not originating from the same \bquark hadron (combinatorial background). The BDT algorithm is implemented with the TMVA toolkit~\cite{TMVA4}.
For the \bpdstdspi process, the selections are trained using simulated events, where all the decays are distributed uniformly across the phase space, as the signal training sample. The background training sample comprises wrong-sign $\Dstarm\Dsm\pip$ and $\Dstarm\Dsp\pim$ combinations in data. In the case of the normalisation channel \bdstds, the simulation sample is used for signal, while $\Dstarm\Dsp$ combinations with invariant masses greater than 5.4\gev are used as the background in the BDT training. The classifier takes as inputs the vertex $\chisq$ for the \B, \Dsp and \Dzb meson candidates, the $\chisq_{\rm IP}$ of the \B, \Dsp, \Dzb candidates and the final-state tracks, the signed significances of the separation of the \Dsp and \Dzb vertices from the \B meson vertex parallel to the beam pipe, and the corrected PID information of the final-state kaon and pion candidates. The decision on the BDT response is made based on retaining a high fraction of signal events while maintaining a relatively pure sample with a purity of 90\%. The estimated purity is calculated as $S/(S + B)$, where $S$ and $B$ are the expected signal and background yields in the $\pm 30\mev$ range around the nominal $B$ meson mass. 

Backgrounds, where one of the final-state tracks is misidentified (misID backgrounds), are suppressed with stricter requirements on the PID variables and the application of vetoes. To mitigate contributions from other double-charm $\bquark$-hadron decays (specifically, the 
\mbox{$\Bz\to\Dstarm\Dp(\to\Km\pip\pip)$} and $\Lb\to\Dstarm\Lc(\to\Km\proton\pip)$ decays), the PID requirements for the \Dsp final-state tracks are tightened for candidates with invariant mass within $\pm 20\mev$ around the \Dp or \Lc mass under the $m(\Km\pip\pip)$ or $m(\Km\proton\pip)$ hypotheses, respectively. Additional misID backgrounds arise from $\Bp\to\Dsm\pim\pip\pip\pip$ and \mbox{$\Bz\to\Dsm\pim\pip\pip$} decays, where double $K\to\pi$ misidentification occurs. Candidates under alternative mass hypotheses lying within a $\pm 30\mev$ interval around the \B meson masses are subsequently rejected.

A significant source of background is due to \Bp and \Bz decays that proceed without the production of the \Dsp meson (non-\Dsp background). Its fraction is calculated by extracting the number of candidates in the invariant-mass peak near the \B mass from the \Dsp sidebands and scaling this number by a factor equal to the ratio between the sizes of the \Dsp signal window and the \Dsp sideband region. The contribution of this background is suppressed by imposing a requirement on the significance of the \Dsp flight distance along the beam direction, denoted as $\Delta z/\sigma(\Delta z)$. To avoid a significant loss of efficiency, a loose cut, $\Delta z/\sigma(\Delta z)>1$, is applied, which reduces the fraction of non-\Dsp decays to 4--6\% (depending on data-taking period) for the \bpdstdspi sample, and to approximately 2\% for the \bdstds sample. The remaining non-\Dsp background is explicitly considered in both the amplitude analysis and the ratio of branching fractions measurement.

After applying the full selection, around 2\% (1\%) of the events include more than one reconstructed \bpdstdspi (\bdstds) candidate. All candidates are retained for subsequent analysis. The extra candidates correspond to the combinatorial background and are treated as such in the fits to extract signal yields. The effect of keeping multiple candidates in the amplitude fit is evaluated by performing an alternative fit keeping one random candidate per event, and is found to be negligible. 

\section{\texorpdfstring{Fits of {\boldmath $\Dstarm\Dsp$} and {\boldmath $\Dstarm\Dsp\pip$} invariant-mass distributions}{Fits of DstarDs and DstarDspi invariant-mass distributions}}
\label{sec:Massfit}

The yields of the signal and normalisation decays are obtained by fitting the $\Dstarm\Dsp\pip$ and $\Dstarm\Dsp$ invariant-mass distributions split into four categories, according to the data-taking period (Run 1 or Run 2) and trigger category (TOS or NotTOS).

The invariant-mass distributions of the $\Bp\to\Dstarm\Dsp\pip$ and $\Bz\to\Dstarm\Dsp$ candidates contain a small admixture from the non-$\Dsp$ background and are referred to as the fully reconstructed $\B$ meson decays. These distributions are parametrised by a sum of a Gaussian function and a double-sided Crystal Ball (DCB) function~\cite{Skwarnicki:1986xj} for each category. The distribution of non-$\Dsp$ events is broader than the signal distribution due to the correlation between the \B and \Ds candidate invariant masses, which is induced by the \Ds mass constraint. It is represented by the convolution of the signal distribution with the rectangular function of a width equal to the \Ds invariant mass range of $\pm 20\mev$. In the fit to the data, the peak positions (shared between the Gaussian and DCB functions), width, and relative magnitude of the Gaussian peak are allowed to float. The parameters describing the tails of the DCB function are fixed to the values obtained from fits to the simulation samples. The $\Dstarm\Dsp$ invariant-mass spectrum includes a small contribution from the $\Bs$ decays~\cite{LHCB-PAPER-2021-006}, with the distribution fixed to be the same as for the $\Bz$ decays, but shifted by the known $\Bs$--$\Bz$ mass difference~\cite{PDG2022}. 

The invariant-mass range below the $\B$ meson mass is populated with candidates from $\B$ meson decays with $\D^{*+}_s$ subsequently decaying as $\D^{*+}_s\to \Dsp\gamma/\piz$, where the photon or the $\piz$ from the decay is not reconstructed. This structure is referred to as the partially reconstructed $\B$ meson decay component. The distribution for the $\Bz\to\Dstarm\D^{*+}_s$ decays is obtained from the corresponding simulated sample and is described by a non-parametric kernel density estimator~\cite{Cranmer:2000du}. The shape of the partially reconstructed $\Bp\to\Dstarm\D_s^{*+}\pip$ decays depends on the unmeasured structure of the decay amplitude and is thus parametrised by an empirical shape, the sum of two Gaussian peaks. 

The combinatorial background is parametrised by an exponential distribution. Its slope is floated in the fit to data. 

\begin{figure}[b]
  \includegraphics[width=0.5\textwidth]{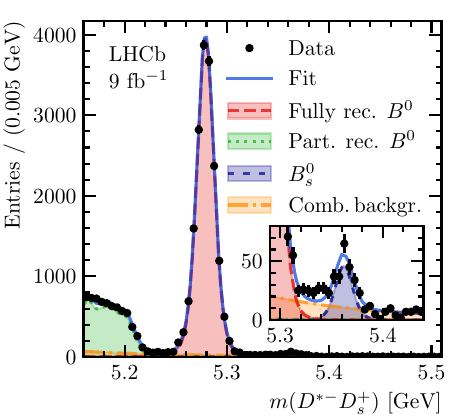}
  \put(-35,170){(a)}
  \includegraphics[width=0.5\textwidth]{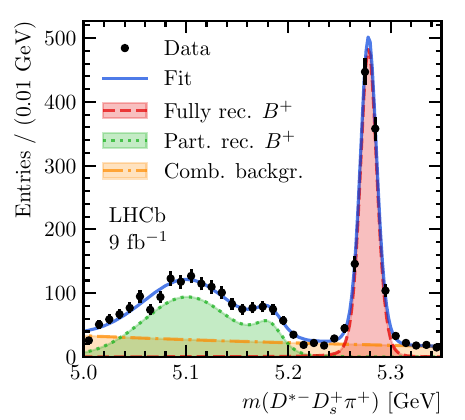}
  \put(-35,170){(b)}
  \caption{Invariant-mass distributions of (a) $\Dstarm\Dsp$ and (b) $\Dstarm\Dsp\pip$ combinations and the results of the fits used to obtain the yields of the $\Bz\to\Dstarm\Dsp$, $\Bp\to\Dstarm\Dsp\pip$ and $\Bp\to\Dstarm\Dssp\pip$ decays. The inset in the plot (a) shows a zoomed region with the contribution of the $\Bs\to\Dstarm\Dsp$ decay component. }
  \label{fig:massfits}
\end{figure}

The invariant-mass distributions of the $\Dstarm\Dsp$ and $\Dstarm\Dsp\pip$ combinations in data and the results of the fits are shown in Fig.~\ref{fig:massfits} for all categories combined. 
The yields of various fit components are given in Table~\ref{tab:dstds_yields} for the $\Dstarm\Dsp$ and in Table~\ref{tab:dstdspi_yields} for the $\Dstarm\Dsp\pip$ combinations. The reported yields are extracted by performing independent fits to the four different categories. For the \bpdstdspi mode, the signal and background yields in the range $|m(\Dstarm\Dsp\pip)-m_{\Bp}|<30\mev$ (``signal box'') are also reported. This range is used to select the candidates for the amplitude fit. 

\begin{table}[t]
    \centering
    \caption{Yields of signal and background components for the $\Dstarm\Dsp$ invariant-mass fit in range 5.15--5.60\gev.}
    \label{tab:dstds_yields}
    \begin{tabular}{l|r|r|r|r}
    & \multicolumn{2}{c|}{Run 1} & 
      \multicolumn{2}{c}{Run 2} \\
      \cmidrule{2-5}
                          & TOS & NotTOS & TOS & NotTOS \\
                          \midrule
      Fully rec. $\Bz$             & $2\,512\pm 53$ & $1\,017\pm 33$ & $9\,720\pm 102$ & $4\,151\pm 67$ \\ 
      Part. rec. $\Bz$           & $1\,101\pm 48$ & $500\pm 24$ & $4\,071\pm 83\phantom{0}$ & $1\,762\pm 63$  \\
      $\Bs$            & $27\pm 7\phantom{0}$    & $14\pm 4\phantom{0}$    & $117\pm 14\phantom{0}$   & $48\pm 9\phantom{0}$    \\
      Comb. backgr.       & $211\pm 54$ & $37\pm 16$ & $994\pm 82\phantom{0}$ & $374\pm 66$  \\
    \end{tabular}
\end{table}

\begin{table}[t]
    \centering
    \caption{Yields of signal and background components for the $\Dstarm\Dsp\pip$ invariant-mass fit in range 4.80--5.60\gev and in the signal box $|m(\Dstarm\Dsp\pip)-m_{\Bp}|<30\mev$.}
    \label{tab:dstdspi_yields}
    \begin{tabular}{l|c|c|c|c}
    & \multicolumn{2}{c|}{Run 1} & 
      \multicolumn{2}{c}{Run 2} \\
      \cmidrule{2-5}
                          & TOS & NotTOS & TOS & NotTOS \\
                          \midrule
      Fully rec. $\Bp$        & $139\pm 14$ & $52\pm 8$ & $598\pm 26$ & $252\pm 17$ \\
      Part. rec. $\Bp$        & $146\pm 16$ & $\phantom{0}60\pm 10$ & $707\pm 37$ & $247\pm 21$  \\
      Comb. backgr.       & $227\pm 20$ & $103\pm 12$ & $1\,129\pm 44\phantom{0\,}$ & $521\pm 28$  \\
      \midrule
      Fully rec. $\Bp$ in signal box       & $132\pm 13$ & $50\pm 8$ & $588\pm 26$ & $249\pm 17$  \\
      Backgrounds in signal box   & $13.5\pm 1.2$ & $\phantom{0}6.4\pm 0.8$ & $63.4\pm 2.5$ & $31.1\pm 1.7$  \\
    \end{tabular}
\end{table}

\begin{figure}
  \includegraphics[width=0.5\textwidth]{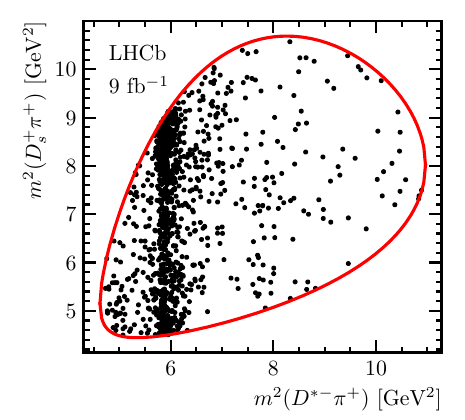}
  \put(-37, 178){(a)}
  \includegraphics[width=0.5\textwidth]{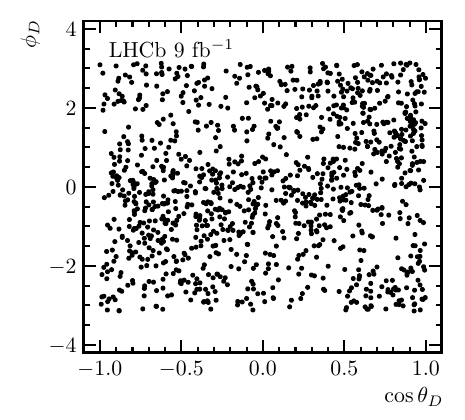}
  \put(-37, 178){(b)}

  \caption{Distribution of $\Bp\to\Dstarm\Dsp\pip$ candidates in data: (a) projection onto the Dalitz plot variables $m^2(\Dstarm\pip)$ and $m^2(\Dsp\pip)$, (b) projection onto angular variables \ensuremath{\cos\theta_D} and \ensuremath{\phi_D}. The solid red line indicates the phase-space boundaries.}
  \label{fig:dalitz}
\end{figure}

Two-dimensional projections of the Dalitz-plot and angular variables for the \mbox{\bpdstdspi} candidates in the signal box are shown in Fig.~\ref{fig:dalitz}. The distributions are not background-subtracted or efficiency-corrected. 

\section{Efficiency}
\label{sec:Efficiency}

The efficiency variations across the four-dimensional phase space are obtained from the simulated signal samples. These samples are generated uniformly in the decay phase space and are analysed using the same reconstruction and selection procedure as for data. The efficiency profiles are computed separately for Run 1 and Run 2 conditions and further split into TOS and NotTOS categories. Prior to performing the density estimation, each simulated event is assigned a weight, derived from control samples in data, to correct for known differences in track reconstruction~\cite{LHCb-DP-2013-002} and hardware trigger~\cite{LHCb-PUB-2014-039} efficiency between data and simulation.

For each category, the efficiency as a function of the phase-space variables is parametrised using an artificial neural network (ANN) density estimator~\cite{Mathad:2019rqj}. In this approach, the weights and biases of a fully connected feed-forward ANN are treated as free parameters in a maximum-likelihood fit to the unbinned simulated data. The first layer of neurons (input layer) is provided with the phase-space variables (\msqdstpi, \msqdspi, \costhd, \phid). To ensure both continuity and periodicity of the estimated density as a function of the \phid angle, the ANN estimator takes both $\cos\phid$ and $\sin\phid$ as inputs instead of the angle itself. The network contains three hidden layers of 40, 80, and 20 neurons, respectively. The output is a single neuron that returns the estimated density.

The optimisation is performed using \tensorflow~\cite{tensorflow2015-whitepaper} with the Adam algorithm~\cite{kingma2014adam}. An L2 regularisation term, calculated as the sum of squared weights multiplied by a tunable parameter $\lambda_2$, is added to the loss function. This term controls overfitting and ensures smoothness in the density function by penalising large neuron weights. The choice of the $\lambda_2$ parameter is driven by the compromise between overfitting for low $\lambda_2$ (which manifests itself as large fluctuations of the fitted density) and systematic bias for high $\lambda_2$ values. The range of valid $\lambda_2$ values is chosen by visually inspecting the projections of the fitted density and their comparison with data, and the middle of this range ($\lambda_2=0.3$) is taken as the $\lambda_2$ parameter for the baseline fit. The upper and lower values of the $\lambda_2$ range are used for the systematic uncertainty evaluation (see Sect.~\ref{sec:uncertainties}). 

\begin{figure}[thb]
  \includegraphics[width=0.48\textwidth]{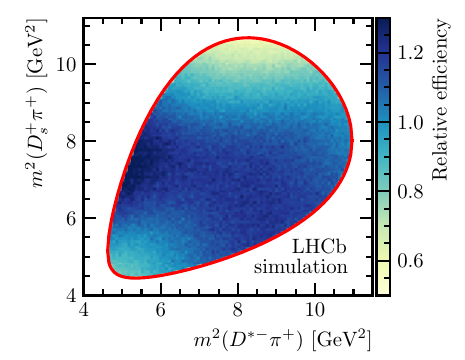}
  \put(-165, 140){(a)}
  \includegraphics[width=0.48\textwidth]{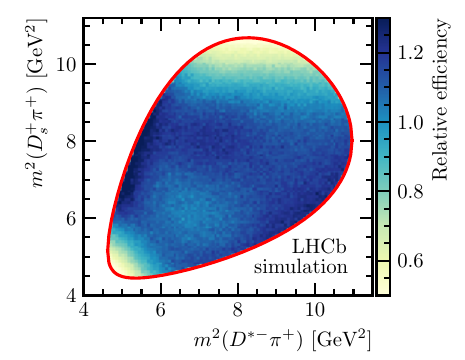}
  \put(-165, 140){(b)}

  \includegraphics[width=0.48\textwidth]{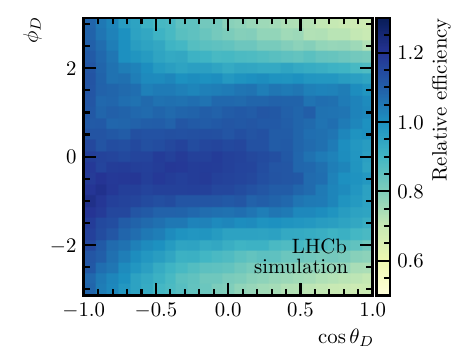}
  \put(-70, 140){(c)}
  \includegraphics[width=0.48\textwidth]{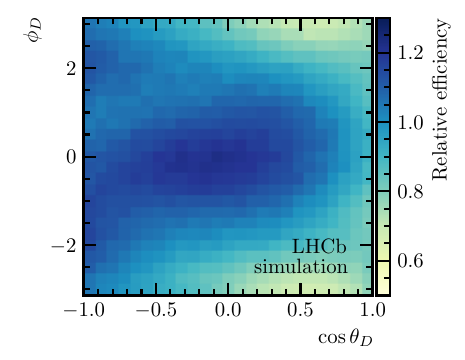}
  \put(-70, 140){(d)}

  \caption{Projections of the $\Bp\to\Dstarm\Dsp\pip$ efficiency profile averaged over trigger categories 
           for (a,c) Run 1 and (b,d) Run 2 onto (a,b) $m^2(\Dstarm\pip)$ and $m^2(\Dsp\pip)$ pair, 
           and (c,d) $\phi_D$ and $\cos\theta_D$ pair. }
  \label{fig:eff}
\end{figure}

The efficiency profile used in the fits is determined as the average of the TOS and NotTOS profiles, weighted according to the ratio of yields of the two categories of events in data. The projections of the resulting efficiency profile onto the Dalitz plot variables \msqdstpi, \msqdspi and the angular variables \costhd, \phid are shown in Fig.~\ref{fig:eff} for Run~1 and Run~2 separately.

\section{Background distributions}
\label{sec:Backgrounds}

The density of the combinatorial background as a function of the phase space variables is derived using candidates in the \Bp upper-mass sideband defined as \mbox{$5.31<m(\Dstarm\Dsp\pip)<5.60\gev$}. The low-mass sideband is excluded as it is dominated by partially reconstructed decays with distinct amplitude structures. 

A five-dimensional probability density, $\mathcal{P}_{\rm bkg}(\mathbf{x}, m(\Dstarm\Dsp\pip))$, is estimated using an ANN density estimator. The inclusion of $m(\Dstarm\Dsp\pip)$ accounts for any potential effect arising from applying the \Bp mass constraint when computing the phase-space variables, $\mathbf{x}$. The resulting $\mathcal{P}_{\rm bkg}(\mathbf{x}, m(\Dstarm\Dsp\pip))$ parametrisation is then used to extrapolate the combinatorial background into the signal region, $\mathcal{P}_{\rm bkg}(\mathbf{x}) \equiv \mathcal{P}_{\rm bkg}(\mathbf{x},m_{\Bp})$~\cite{Mathad:2019rqj}. 

The density estimation is performed on the combined Run 1 and Run 2 \Bp data upper-mass sideband. The architecture of the ANN estimator aligns with that of the efficiency density estimation, except that the input layer incorporates an additional neuron to capture the $m(\Dstarm\Dsp\pip)$ variable. The $\lambda_2$ regularisation parameter is chosen similarly to the case of the efficiency shape (Sect.~\ref{sec:Efficiency}) and set to 0.2. An extra L2 regularisation term with $\lambda_2 = 30$ is introduced for the neuron weights of the first hidden layer corresponding to the $m(\Dstarm\Dsp\pip)$ input to ensure smooth dependence of the distribution in phase-space variables on the $m(\Dstarm\Dsp\pip)$ invariant mass.

\begin{figure}[htb]
  \centering
  \includegraphics[width=0.32\textwidth]{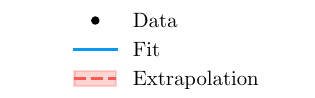}

  \includegraphics[width=0.32\textwidth]{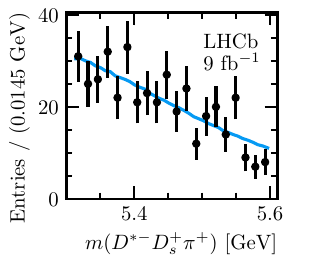}
  \put(-105, 38){(a)}
  \includegraphics[width=0.32\textwidth]{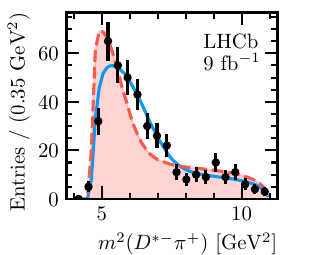}
  \put(-42, 48){(b)}
  \includegraphics[width=0.32\textwidth]{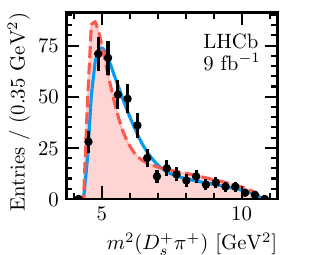}
  \put(-42, 48){(c)}

  \includegraphics[width=0.32\textwidth]{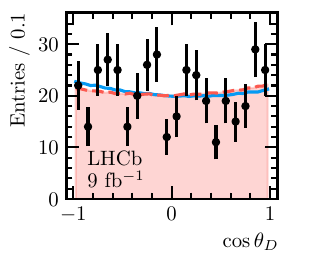}
  \put(-42, 37){(d)}
  \includegraphics[width=0.32\textwidth]{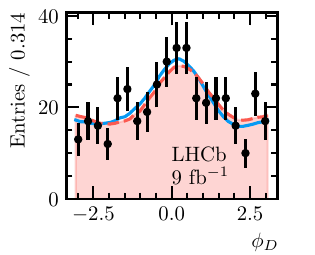}
  \put(-42, 95){(e)}
  \includegraphics[width=0.32\textwidth]{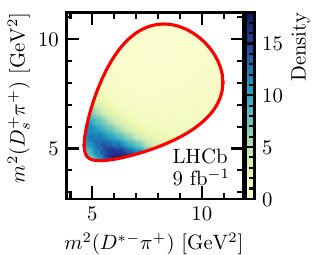}
  \put(-107, 94){(f)}

  \includegraphics[width=0.32\textwidth]{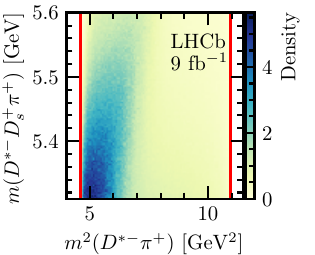}
  \put(-58, 37){(g)}
  \includegraphics[width=0.32\textwidth]{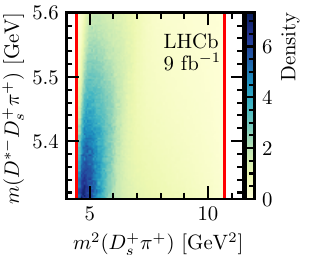}
  \put(-60, 37){(h)}
  \includegraphics[width=0.32\textwidth]{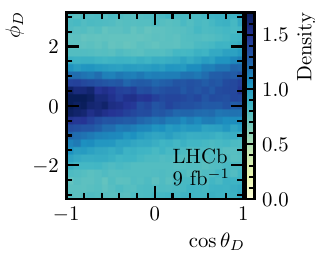}
  \put(-52, 95){(i)}

  \caption{(a--e) One-dimensional projections of the distribution of $\Bp\to\Dstarm\Dsp\pip$ candidates in the invariant mass range $5.31<m(\Dstarm\Dsp\pip)<5.60$\gev (upper $\Bp$ invariant-mass sideband), results of its density estimated with ANN, and, except for (a), results of the extrapolation of its density to the signal region with $m(\Dstarm\Dsp\pip) = m_{\Bp}$. (f--i) Two-dimensional projections of the estimated density of $\Bp\to\Dstarm\Dsp\pip$ candidates in the upper $\Bp$ mass sideband. }
  \label{fig:comb_backgr}
\end{figure}

Projections of the estimated five-dimensional background density and the distribution of combinatorial background candidates in the upper mass sideband are shown in Fig.~\ref{fig:comb_backgr}. The projections of the phase-space variable distribution extrapolated to the signal region, $m(\Dstarm\Dsp\pip) = m_{\Bp}$, are also presented.

After the final selection, the \bpdstdspi sample is still contaminated by non-\Dsp background (see Sect.~\ref{sec:Selection}), which has to be explicitly included in the amplitude fit. This is done by modifying the total PDF as defined in Eq.~\ref{eq:totalpdf} to incorporate the parametrisation of the non-\Dsp background:
\begin{equation}
    \mathcal{P}_{\rm tot}(\mathbf{x}) = \left(\left\vert\mathcal{A}(\mathbf{x})\right\vert^{2}\epsilon(\mathbf{x})\frac{1-f_{{\rm non}\mhyphen\Dsp}}{\mathcal{N}_{\rm sig}}+\mathcal{P}_{{\rm non}\mhyphen\Dsp}\left(\mathbf{x}\right)\frac{f_{{\rm non}\mhyphen\Dsp}}{\mathcal{N}_{{\rm non}\mhyphen\Dsp}}\right)\left(1-f_{\rm bkg}\right)+\mathcal{P}_{\rm bkg}(\mathbf{x})\frac{f_{\rm bkg}}{\mathcal{N}_{\rm bkg}}. 
    \label{eq:totalpdfnonDs}
\end{equation}
Here the $\mathcal{P}_{{\rm non}\mhyphen\Dsp}(\mathbf{x})$ represents the density function of the non-\Dsp background component, also parametrised using an ANN estimator with $\mathcal{N}_{{\rm non}\mhyphen\Dsp}$ as the normalisation term, and $f_{{\rm non}\mhyphen\Dsp}$ denotes its estimated fractional contribution.

\begin{figure}[htb]
  \includegraphics[width=0.32\textwidth]{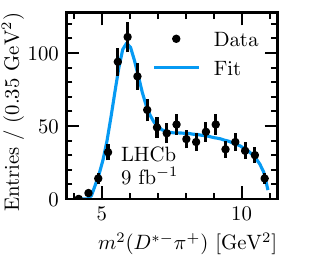}
  \put(-107, 94){(a)}
  \includegraphics[width=0.32\textwidth]{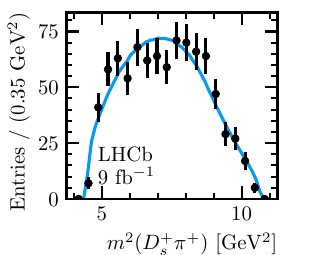}
  \put(-42, 94){(b)}
  \includegraphics[width=0.32\textwidth]{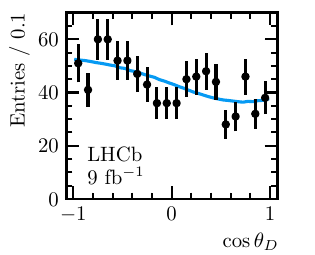}
  \put(-42, 37){(c)}

  \includegraphics[width=0.32\textwidth]{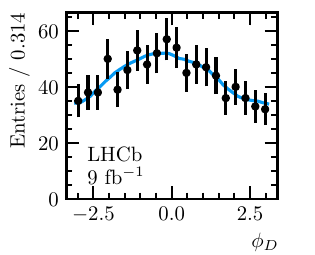}
  \put(-42, 37){(d)}
  \includegraphics[width=0.32\textwidth]{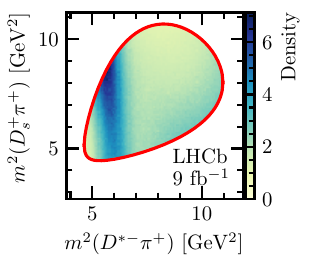}
  \put(-107, 94){(e)}
  \includegraphics[width=0.32\textwidth]{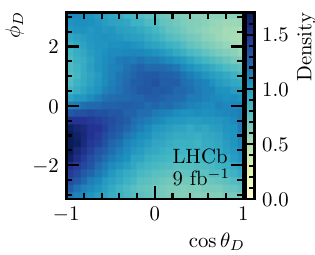}
  \put(-54, 94){(f)}

  \caption{(a--d) One-dimensional projections of the distribution of $\Bp\to\Dstarm\Dsp\pip$ candidates in the \Dsp sideband region and the results of its density estimation using ANN. (e, f) Two-dimensional projections of the estimated non-$\Dsp$ background density. }
  \label{fig:nonds_backgr}
\end{figure}

The shape of the non-\Dsp background is obtained from the combination of Run 1 and Run 2 events in the \Dsp sidebands and requiring $\Delta z/\sigma(\Delta z)<2$. The distributions and the results of the density estimation are presented in Fig.~\ref{fig:nonds_backgr}. The same ANN architecture as in the efficiency density estimation is used, with the $\lambda_2$ regularisation parameter set to 0.18, following the same approach as described in Sect.~\ref{sec:Efficiency}. 

\section{Amplitude analysis}
\label{sec:Amplitude}

Various amplitude models are employed to fit the \bpdstdspi data. The Run 1 and Run 2 samples are fitted simultaneously. The baseline fit exclusively incorporates resonance activity in the $\Dstarm\pip$ channel. Alternative fits, exploring exotic contributions in the $\Dsp\pip$ and $\Dstarm\Dsp$ channels, are also examined. This is motivated by the data and the amplitude analysis of the $\Bp\to\Dm\Dsp\pip$ decays, where an exotic state in the $\Dsp\pip$ channel has been identified~\cite{LHCb-PAPER-2022-026}. 

\subsection{Baseline fit}
\label{subsec:baseline_fit}
In the baseline model, the parametrisation of excited charm-meson resonances decaying to the $\Dstarm\pip$ final state employs Breit--Wigner line shapes. The considered states are listed in Table~\ref{tab:excitedDPDG}, with the default inclusion of $D_{1}(2420)$, $D_{1}(2430)$, and $D_{2}^{*}(2460)$ states. 
The fits with all possible combinations including or not the remaining $\D_0(2550)$, $\D^*_1(2600)$, $\D_2(2740)$, and $\D^*_3(2750)$ states are performed. Masses and widths are fixed to the values reported in Ref.~\cite{PDG2022}. The data are fitted with the complex $LS$ couplings for each amplitude component treated as floating parameters, except for the $D_{1}(2420)$ D-wave amplitude, which is taken as a reference and set to unity. The decision to include additional resonances is contingent on the variation of the fit likelihood and the corresponding fit fraction.

\begin{table}
\centering
\caption{\small Resonances considered in the baseline amplitude model and their parameters.}
\label{tab:excitedDPDG}
\begin{tabular}{c|clc}
	Resonance & $J^{P}$ & Mass [MeV] & Width [MeV]\\
	\midrule
	$D_{1}(2420)$ &$1^{+}$ & $2\,422.1\pm0.6$ & $31.3\pm1.9$\\
	$D_{1}(2430)$ &$1^{+}$ & $2\,412\phantom{.0}\pm9\phantom{.0}$ & $314\pm29$\\
        $D_{2}^{*}(2460)$ & $2^{+}$ & $2\,461.1\;{}^{+0.7}_{-0.8}$ & $47.3\pm0.8$\\
	\midrule
	$D_{0}(2550)$ &$0^{-}$ & $2\,549\phantom{.0}\pm19\phantom{.}$ & $165\pm24$\\
	$D_{1}^{*}(2600)$ &$1^{-}$ & $2\,627\phantom{.0}\pm10\phantom{.}$ & $141\pm23$\\
	$D_{2}(2740)$ &$2^{-}$ & $2\,747\phantom{.0}\pm6\phantom{.0}$ & $\phantom{0}88\pm19$\\
	$D_{3}^{*}(2750)$ &$3^{-}$ & $2\,763.1\pm 3.2$ & $66\pm 5$\\
\end{tabular}
\end{table}

Considering all the resonant compositions of the fit model, the inclusion of the $D_{0}(2550)$, $D_{1}^{*}(2600)$, and $D_{2}(2740)$ states improves the fit likelihood, yielding fit fractions larger than 1\%. Conversely, the $D_{3}^{*}(2750)$ fit fraction is below 1\%, and the NLL difference stands at approximately three units for four additional degrees of freedom. Consequently, the $D_{3}^{*}(2750)$ state is excluded from the baseline model. Estimations of the significances for the remaining three resonances are derived from the NLL difference for fits with and without the corresponding state, resulting in estimated significances of $6.5\sigma$ for $D_{0}(2550)$, $6.8\sigma$ for $D_{1}^{}(2600)$, and $4.6\sigma$ for $D_{2}(2740)$ states.

\begin{figure}
  \includegraphics[width=0.48\textwidth]{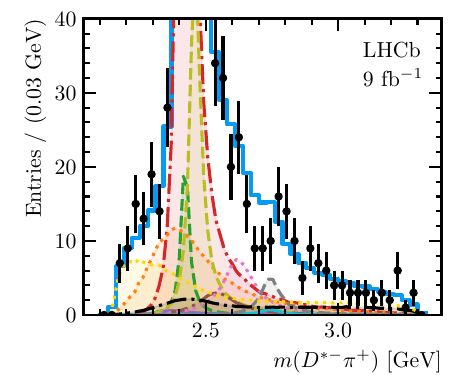}
  \put(-175, 150){(a)}
  \includegraphics[width=0.48\textwidth]{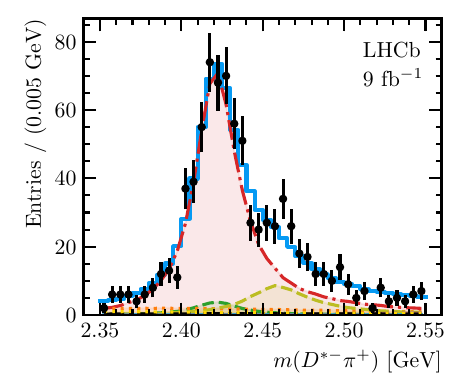}
  \put(-175, 150){(b)}

  \includegraphics[width=0.48\textwidth]{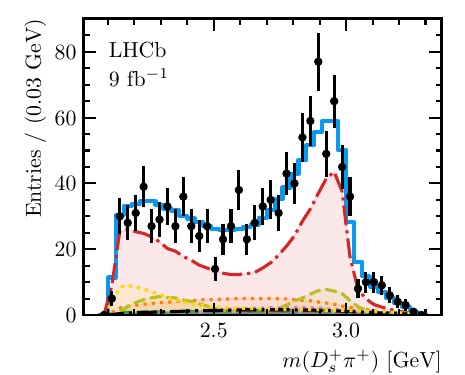}
  \put(-40, 150){(c)}
  \includegraphics[width=0.48\textwidth]{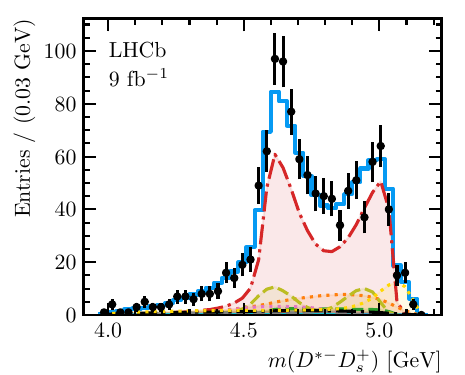}
  \put(-40, 150){(d)}

  \includegraphics[width=0.48\textwidth]{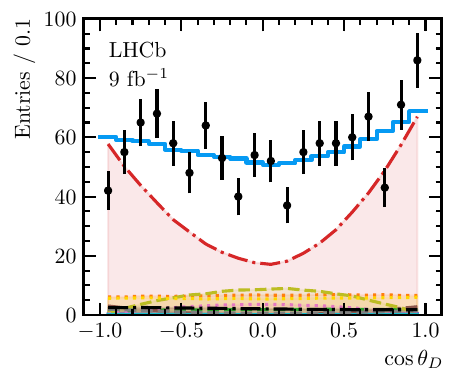}
  \put(-50, 150){(e)}
  \includegraphics[width=0.48\textwidth]{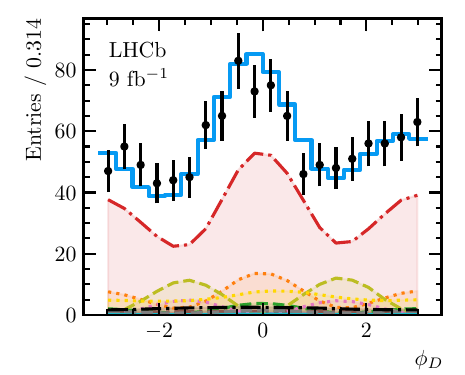}
  \put(-40, 150){(f)}

  \includegraphics[width=\textwidth]{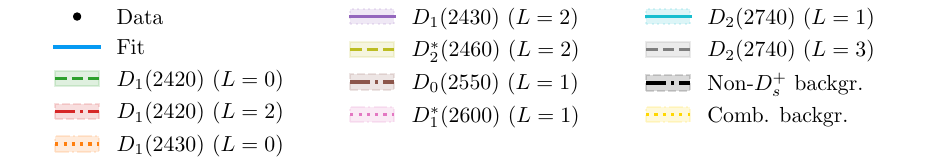}

  \caption{Results of the fit of the $\Bp\to\Dstarm\Dsp\pip$ distribution with the baseline model. Figures (a) and (b) show the $m(\Dstarm\pip)$ projection, with (a) zoomed in to illustrate the contributions from all the resonances while (b) shows the projection near the $D_1(2420)$ resonance. The $m(\Dsp\pip)$, $m(\Dstarm\Dsp)$, $\cos\theta_{D}$ and $\phi_{D}$ projections are shown in (c), (d), (e) and (f), respectively.}
  \label{fig:fit}
\end{figure}

\begin{figure}
  \includegraphics[width=0.48\textwidth]{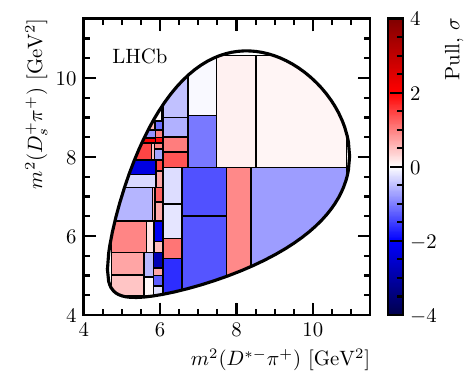}
  \put(-70, 150){(a)}
  \includegraphics[width=0.48\textwidth]{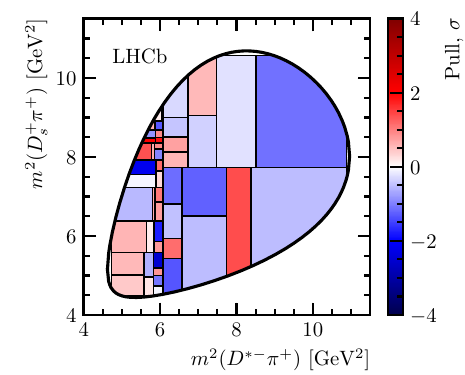}
  \put(-70, 150){(b)}

  \caption{Pulls of the $\Bp\to\Dstarm\Dsp\pip$ Dalitz-plot distribution for the fits with (a) the baseline model and (b) the model including 
           $\Dsp\pip$ components. }
  \label{fig:pulls}
\end{figure}

The results of the baseline fit, along with the alternative fits discussed in Sect.~\ref{subsec:alt_fits}, are presented in Appendix~\ref{apdx:resultfits}. The invariant-mass and angular projections of the baseline fit result are presented in Fig.~\ref{fig:fit}. To assess the fit quality, a $\chisq/{\rm ndof}$ value is computed from the two-dimensional distribution $(m^{2}(\Dstarm\pip), m^{2}(\Dsp\pip))$. Adaptive binning is constructed such that each bin contains a minimum of 25 entries. The resulting fit quality is $\chisq/{\rm ndof}=58.6/46$, where the effective number of degrees of freedom is obtained from simulated pseudoexperiments. Figure~\ref{fig:pulls}~(a) illustrates the pulls in bins of the Dalitz plot, using the same binning employed for the $\chi^2/{\rm ndof}$ calculation.

The $m(\Dsp\pip)$ projection reveals a certain excess of data over the baseline fit around 2.9\gev. The pull distribution indicates that this enhancement is particularly significant for invariant masses $m^2(\Dstarm\pip)<6\gev^2$, inconsistent with attributing it to an additional resonant spin-zero state in the $\Dsp\pip$ channel with the mass around 2.9\gev. Nevertheless, additional studies are conducted in an attempt to improve the description of the data, considering exotic $\Dsp\pip$ contributions. Results from those fits are discussed in Sect.~\ref{subsec:exo_fits}.

Ensembles of pseudoexperiments, where the baseline model is used both to generate and to fit samples of the same size as in the data, are used to evaluate the statistical uncertainties on the fit fractions and check for systematic biases due to the fitting procedure as discussed in Sect.~\ref{sec:uncertainties}.

Table~\ref{tab:ff_results} provides the obtained fit fractions for the components of the baseline amplitude model and the phase differences between the amplitude components and the reference $D_1(2420)$ D-wave amplitude. The reported values include both statistical and systematic uncertainties. The assessment of systematic uncertainties is detailed in Sect.~\ref{sec:uncertainties}. Notably, all components of the baseline model, except for the $1^+$ resonances $D_1(2420)$ and $D_1(2430)$, have different quantum numbers. Therefore, only the interference terms between these two components are different from zero after integration over the phase space.

Since the \bpdstdspi decay amplitude is dominated by the favoured $b\to c$ transition, \CP-violating effects in it are anticipated to be minimal. Independent fits of the baseline model to the \Bp and \Bm samples are performed and exhibit statistical agreement. Accounting for the correlation of the fit parameters, the $p$-value of the agreement between them is 19\%, equivalent to a difference of 1.3 standard deviations.

The consistency between the Run 1 and Run 2 datasets is tested by conducting separate fits to the data. The fit utilises the selection efficiencies obtained separately for Run~1 and Run~2 simulated samples, while the background parametrisation is taken to be common, identical to the one used in the baseline fit. The parameters of the fit exhibit good statistical agreement for the two independent samples. 

\begin{table}
\centering
\caption{Fit fractions (in \%) for the components of the $\Bp\to\Dstarm\Dsp\pip$ amplitude. The first uncertainty is statistical, 
the second is systematic, and the third is the uncertainty related to the amplitude model. }
\label{tab:ff_results}
\begin{tabular}{l|rr}
Component & Fit fraction [\%] & Phase [rad] \\
\midrule
$D_1(2420)$ S-wave  &  $3.8\pm 1.7\pm 0.8^{+1.3}_{-0.1}$ & $-1.96\pm 0.16\pm 0.10^{+0.17}_{-0.05}$ \\
$D_1(2420)$ D-wave  & $71.0\pm 4.4\pm 4.6^{+0.0}_{-6.0}$ & 0 (fixed) \\
$D_1(2430)$ S-wave  & $14.2\pm 2.5\pm 2.4^{+3.1}_{-2.0}$ & $+0.14\pm 0.11\pm 0.13^{+0.06}_{-0.18}$ \\
$D_1(2430)$ D-wave  &  $0.5\pm 0.9\pm 1.5^{+0.2}_{-0.5}$ & $-2.99\pm 0.42\pm 0.84^{+0.23}_{-0.55}$ \\
$D^*_2(2460)$         & $11.7\pm 1.4\pm 0.8^{+0.0}_{-0.7}$ & $+3.14\pm 0.11\pm 0.14^{+0.05}_{-0.04}$ \\
$D_0(2550)$           &  $2.3\pm 0.8\pm 0.7^{+0.3}_{-1.7}$ & $-2.24\pm 0.21\pm 0.26^{+0.05}_{-0.25}$ \\
$D^*_1(2600)$         &  $4.8\pm 1.0\pm 0.9^{+1.1}_{-2.0}$ & $+0.32\pm 0.16\pm 0.16^{+0.37}_{-0.01}$ \\
$D_2(2740)$ P-wave  &  $0.4\pm 0.4\pm 0.2^{+0.1}_{-0.1}$ & $-0.02\pm 0.56\pm 0.32^{+0.16}_{-0.59}$ \\
$D_2(2740)$ F-wave  &  $2.3\pm 0.7\pm 0.9^{+0.4}_{-0.1}$ & $-0.09\pm 0.27\pm 0.21^{+0.08}_{-0.23}$ \\
\midrule
Sum of fit fractions  &  $111.0\pm 5.2\pm 4.2\phantom{{}^{+0.0}}$ & \\
\end{tabular}
\end{table}

\subsection{\texorpdfstring{Alternative parametrisations of the {\boldmath $\Dstarm\pip$} amplitude}{Alternative parametrisations of the Dstarpi amplitude}}
\label{subsec:alt_fits}

The use of the Breit--Wigner line shape may be less suitable for amplitudes involving broad resonances or overlapping contributions with the same quantum numbers, such as the case of the broad $D_{1}(2430)$ state overlapping with $D_1(2420)$ in the $1^+$ S-wave of this analysis. Consequently, an alternative strategy is explored, involving a quasi-model-independent description (QMI) of the broad $1^{+}$ S-wave amplitude. In this approach, the amplitude is represented by a complex-valued cubic spline, while all other amplitudes are parametrised by Breit--Wigner functions with masses and widths fixed to the values given in Table~\ref{tab:excitedDPDG}. A spline with six knots is employed, with the first and last knots set to the kinematic limits ($m_{\Dstarm}+m_{\pip})\simeq 2.15\gev$ and $(m_{\Bp}-m_{\Dsp})\simeq 3.31\gev$, respectively. Internal knots are fixed at the masses 2.2, 2.3, 2.5, and $3.0\gev$, and the corresponding six complex coefficients are floated in the fit. 

Comparing the QMI approach with the Breit--Wigner description of the $D_{1}(2430)$ state, the NLL difference is only $-\Delta\ln\mathcal{L}=-4.0$ (see Table~\ref{tab:aman_fit_results} in Appendix~\ref{apdx:resultfits} under the column ``QMI''), while the number of floating parameters is $N_{\rm par}=24$, compared to $N_{\rm par}=16$ in the Breit--Wigner model. Thus, it is concluded that the model-independent description does not yield a significant improvement relative to the Breit--Wigner description given the available statistics.

As an additional test, the mass and width of the Breit--Wigner line shape characterising the $D_{1}(2430)$ state are allowed to vary in the amplitude fit (the model that appears as ``Floated $\D_1(2430)$'' in Table~\ref{tab:aman_fit_results}). However, this alternative fit yields only a marginal improvement in the NLL compared to the fit where the $D_1(2430)$ parameters are fixed to the values given in Ref.~\cite{PDG2022}. The resulting values for the $D_1(2430)$  mass and width are $2.378\pm 0.025$\gev and $0.24\pm 0.06$\gev, respectively, consistent with the values reported in Ref.~\cite{PDG2022}.

In the baseline model, the S- and D-wave decays of the $1^+$ $D_1(2420)$ and $D_1(2430)$ states are considered independent. Therefore, the mixing of the two $1^+$ states is implicitly accounted for. As an alternative parametrisation of the $1^+$ $\Dstarm\pip$ amplitude, the model with the mixing angle $\omega$ and the phase $\psi$, identical to the one used in the \mbox{$\Bp\to\Dstarm\pip\pip$} amplitude analysis~\cite{LHCB-PAPER-2019-027} is used. The mixing parameters are determined to be \mbox{$\omega = -0.054\pm 0.018$} and \mbox{$\psi = 1.23\pm 0.72$}, where the uncertainties are statistical only, consistent with those obtained in Ref.~\cite{LHCB-PAPER-2019-027}. However, since the phases of all the amplitude components in this model are strongly correlated with the phase $\psi$, and given that $\psi$ is not determined precisely, this model is not included in the list of variations to assess the model uncertainty. 

Other tests involve the inclusion of additional broad nonresonant contributions in different $\Dstarm\pip$ partial waves. The line shapes are parametrised with the exponential functions multiplied by the orbital barrier factors for the amplitudes with orbital momentum $L>0$. For the spin-parity combinations where two partial waves are possible, both partial waves are allowed in this fit with floated couplings, but the exponential slope is constrained to be equal for both waves. 

The only partial wave where the addition of an exponential nonresonant amplitude results in a significant improvement of the fit is the $J^P=1^-$ wave. Results of this fit are presented in Table~\ref{tab:aman_fit_results} under the ``NR $D^*\pi$ $1^{-}$'' column. Nevertheless, the addition of the nonresonant $1^-$ component does not result in a substantial change in any of the couplings or the resonance fit fractions, except for the $D_1^*(2600)$ state which has the same quantum numbers, $1^-$, as the added nonresonant amplitude. The fit fraction for this state reduces from ($4.9\pm 1.1$)\% to ($3.0\pm 1.3$)\%, where the uncertainties are statistical only.

\subsection{Fits with exotic contributions}
\label{subsec:exo_fits}

Various fits incorporating $\Dsp\pip$ amplitudes are examined, prompted by the discovery of the $T_{c\bar{s}0}^{\ast}(2900)^{++,0}$ states~in $\Bp\to\Dm\Dsp\pip$ and $\Bz\to\Dzb\Dsp\pim$ decays~\cite{LHCb-PAPER-2022-026, LHCb-PAPER-2022-027}. All of them include the same resonances as in the baseline model, with one or two additional resonant (Breit--Wigner) or nonresonant (exponential) amplitudes. The fits do not reveal evidence of a scalar $\Dsp\pip$ state with the parameters fixed to those of the $\tcsbar$ state~\cite{LHCb-PAPER-2022-026}. A fit with the scalar state and floated parameters results in a very narrow state and only a marginal improvement in the fit likelihood. Introducing a vector state improves the fit but yields a relatively broad amplitude with $\Gamma\simeq 0.4$\gev.  A tensor state with floating parameters produces a very broad peak with the width reaching the upper limit of $1\gev$ with smaller significance than for the vector one. 

Repeating similar fits but with exponential nonresonant amplitudes instead of the Breit--Wigner line shape, the vector and tensor nonresonant amplitudes result in an improved fit likelihood. However, the vector nonresonant model, which yields the best fit, generates a rising amplitude with the exponential parameter $\alpha<0$, which is physically implausible. Incorporating a scalar $\Dsp\pip$ state with the mass and width fixed to that of the $\tcsbar$ state into the amplitude with the additional vector $\Dsp\pip$ contribution leads to further improvement of the fit. Floating the mass and width of the scalar state yields the parameters statistically consistent with those reported in Ref.~\cite{LHCb-PAPER-2022-026}.

\begin{figure}
  \includegraphics[width=0.48\textwidth]{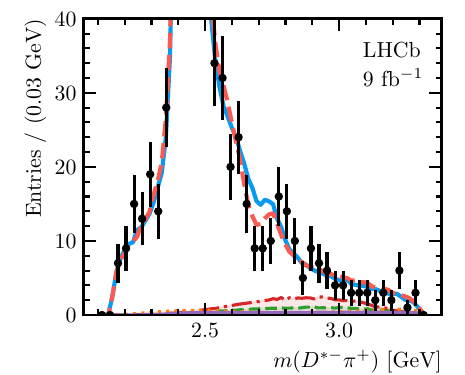}
  \put(-175, 150){(a)}
  \includegraphics[width=0.48\textwidth]{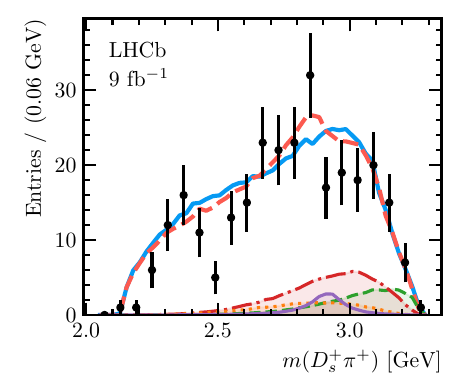}
  \put(-40, 150){(b)}

  \includegraphics[width=0.48\textwidth]{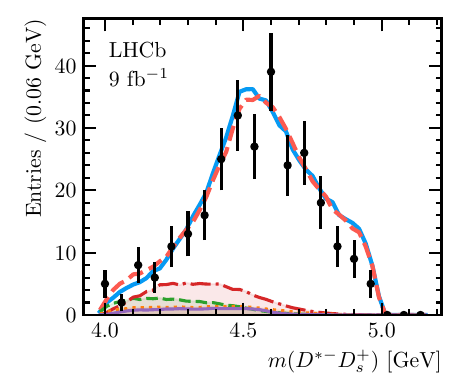}
  \put(-40, 150){(c)}
  \includegraphics[width=0.48\textwidth]{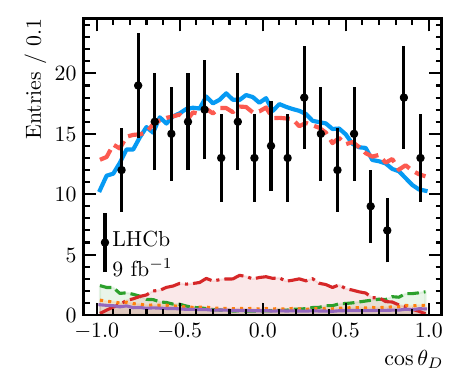}
  \put(-50, 150){(d)}

  \includegraphics[width=0.48\textwidth]{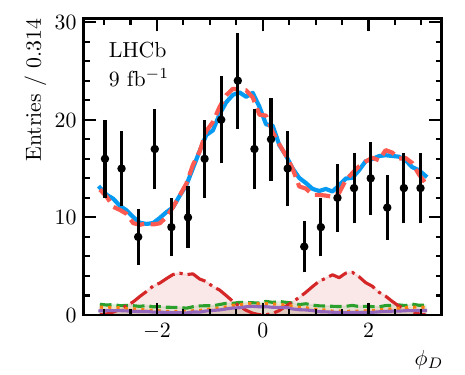}
  \put(-40, 150){(e)}
  \raisebox{0.10\textwidth}{
  \includegraphics[width=0.48\textwidth]{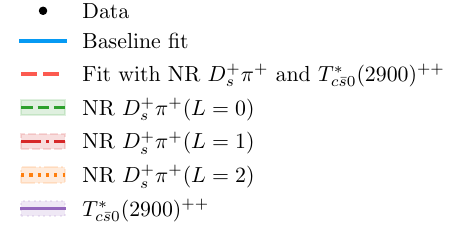}
  }

  \caption{Comparison of the fit results of the $\Bp\to\Dstarm\Dsp\pip$ distribution with the baseline model and with the model that includes $\Dsp\pip$ components. The $m(\Dstarm\pip)$, $m(\Dsp\pip)$, $m(\Dstarm\Dsp)$, $\cos\theta_{D}$ and $\phi_{D}$ projections are shown in (a), (b), (c), (d) and (e), respectively. The requirement $m(\Dstarm\pip)>2.5\gev$ is applied to the distributions shown in plots (b)--(e). }
  \label{fig:fit_tcs}
\end{figure}

Among all the models explored with the additional exotic $\Dsp\pip$ amplitude, the one featuring the nonresonant vector and resonant scalar state, with the parameters fixed to those of the $\tcsbar$ state from the $\Bp\to \Dm\Dsp\pip$ analysis, is selected for further evaluation. The values of the fitted parameters are reported in Table~\ref{tab:aman_fit_results} under the column ``NR, \tcsbar''. Figure~\ref{fig:fit_tcs} illustrates the difference between the baseline model and the model with $\Dsp\pip$ components, emphasised by the requirement $m(\Dstarm\pip)>2.5\gev$ applied to all projections except for $m(\Dstarm\pip)$ itself, to eliminate the dominant $\D^{**}$ states. The fit quality is \mbox{$\chisq/{\rm ndof}=51.3/46$} employing the same binning scheme used to evaluate the fit quality of the baseline model. Figure~\ref{fig:pulls}~(b) displays the pulls in bins of the Dalitz plot using this model. 

The significance and fit fraction of the doubly charged scalar state in the $\Dsp\pip$ channel depend significantly on the model. In the fit where this state is added to the baseline model, its fitted contribution is negligible, with a fit fraction of approximately 0.1\%. However, when incorporated into the model with the $1^-$ $\Dsp\pip$ amplitude, the fit fraction of this state increases to $1.2\pm 0.8$\%, and its statistical significance reaches 2.6$\sigma$. While the conclusion suggests no compelling evidence for the contribution of this state in the $\Dsp\pip$ channel, an upper limit on its fit fraction is established. Using the fitted value of $1.2$\%, along with a statistical uncertainty of 0.8\% obtained from pseudoexperiments, and a systematic uncertainty of 0.5\%, the upper limit is set at 2.3\% (at 90\% CL), or 2.7\% (at 95\% CL) assuming both the statistical and systematic uncertainties to be Gaussian-distributed. It is consistent with the fit fraction for the $\tcsbar\to\Dsp\pip$ contribution of $(2.25\pm 0.67\pm 0.77)\%$ measured in the analysis of $\Bp\to\Dm\Dsp\pip$ decays~\cite{LHCb-PAPER-2022-027}. 

For completeness, fits incorporating exotic contributions in the $\Dstarm\Dsp$ channel are conducted using the baseline model, extended with various amplitudes. None of the models with a single $\Dstarm\Dsp$ amplitude shows a substantial improvement over fits with additional nonresonant states in the $\Dstarm\pip$ or $\Dsp\pip$ channels. The best fit is achieved with $J^P=2^+$ amplitudes, resulting in a very narrow state near the upper kinematic boundary for resonant amplitudes or an amplitude rising with $\Dstarm\Dsp$ mass for nonresonant ones. When incorporating combinations of nonresonant shapes in the $\Dstarm\Dsp$ and $\Dstarm\pip$ channels, the best fits are achieved by models featuring two tensors or two vectors of opposite parity, both resulting in an exotic amplitude rising with the $\Dstarm\Dsp$ mass. It is concluded that none of these fits provides a physical description of the amplitude, and therefore are not considered as part of the model uncertainty.

\subsection{Systematic uncertainties}
\label{sec:uncertainties}

The analysis considers various sources of systematic uncertainties affecting the fit parameters, including $LS$ amplitudes, phases, and fit fractions. Tables~\ref{tab:unc1} and~\ref{tab:unc2} in Appendix~\ref{apdx:unc} present the systematic uncertainties for the baseline amplitude model. 

Systematic uncertainties associated with the efficiency and geometric acceptance description are due to the finite size of simulation samples used for their parametrisation, as well as due to the calibration applied to correct the simulation PID responses and hardware trigger efficiencies. Sample size uncertainty is evaluated by fitting the data employing different efficiency (acceptance) profiles obtained by bootstrapping the original sample, with the standard deviation assigned as the associated uncertainty. The PID correction uncertainty is estimated using an alternative efficiency shape after regenerating the PID responses in simulation with a modified kernel width used for density estimation~\cite{Aaij:2018vrk}. The trigger efficiency correction uncertainty is evaluated by computing efficiencies from simulation samples without trigger efficiency correction. In both instances, systematic uncertainties are established by comparing the results to those of the baseline fit. 

Additional systematic uncertainties arise from the choice of the structure of hidden ANN layers and the regularisation parameter $\lambda_{2}$ in the efficiency (acceptance) density estimation. Alternative parametrisations are obtained with various configurations of ANN hidden layers and by altering the $\lambda_{2}$ regularisation parameter within the range 0.20--0.45. Fits to the data are performed for each alternative efficiency shape, and the largest difference observed compared to the baseline fit is taken as the corresponding uncertainty.

Similar sources of systematic uncertainty arise from the description of the combinatorial (non-\Dsp) background: the finite size of the combinatorial (non-\Dsp) background sample, the choice of the hidden ANN layer structure, and the regularisation parameters used in density estimation. These are assessed employing the same strategy as for the efficiency. Additionally, the uncertainty related to the fraction of combinatorial (non-\Dsp) background in the signal region is evaluated by varying the estimated contribution within its statistical uncertainties.

A sample of $10^6$ events distributed uniformly over the four-dimensional phase space of the decay is used to normalise the PDFs in the amplitude fit. To evaluate the uncertainty due to the precision of PDF normalisation, alternative normalisation samples are generated by bootstrapping the original one. Fits to the data are then carried out using these samples, and the standard deviation is taken as the associated uncertainty.

Fixed parameters in the baseline model are the masses and widths of Breit--Wigner amplitudes, along with the radial parameter in the Blatt--Weisskopf form factors $F_{R}$ and $F_{\B}$ set to $d=4.5$\gev. The associated systematic uncertainty linked to the Blatt--Weisskopf radius is evaluated by varying the radial parameters between 3 and 6\gev. The deviation from the baseline result is then considered as the associated systematic uncertainty. The uncertainty on resonance parameters is evaluated by varying each parameter by its total uncertainty. Positive and negative variations are averaged and added in quadrature to estimate the overall uncertainty.

The inherent biases of the fitting procedure are assessed by fitting ensembles of pseudoexperiments. The same ensembles as those used for the calculation of the statistical uncertainties on the fit fraction are considered. For each parameter, the difference between the baseline fit results and the central values obtained from the pseudoexperiments is considered as the associated systematic uncertainty.

For the assessment of model uncertainty, four alternative models are considered, detailed in Sects.~\ref{subsec:alt_fits} and \ref{subsec:exo_fits}, with the parameters reported in Appendix~\ref{apdx:resultfits}. Asymmetric uncertainties on couplings and fit fractions are assigned as the maximum positive and negative deviations from the baseline model fit. 

\section{Branching fraction measurement}
\label{sec:Branching}
The branching fraction of the $\Bp\to\Dstarm\Dsp\pip$ decay is measured relative to the branching fraction of the $\Bz\to\Dstarm\Dsp$ decay as 
\begin{equation}
	\mathcal{R} =  \frac{\BR(\bpdstdspi)}{\BR(\bdstds)} = \frac{N^{\rm corr}_{\Bp}}{N^{\rm corr}_{\Bz}},
\end{equation}
where $N^{\rm corr}_{\Bp}$ ($N^{\rm corr}_{\Bz}$) is the efficiency-corrected yield of the $\Bp$ ($\Bz$) decay. The analysis measures this ratio from a fit to the \bpdstdspi and \bdstds invariant-mass distributions. The fit is performed simultaneously in the Run 1 (TOS and NotTOS) and Run 2 (TOS and NotTOS) categories, where the uncorrected yields, $N^{i}_{\Bp}$ and $N^{i}_{\Bz}$ (after subtracting the non-\Dsp contributions) in each category, $i$, are related through $\mathcal{R}$ according to 
\begin{equation}
	N^{i}_{\Bp}\;\epsilon^{i}_{\Bp} = \mathcal{R}\;N^{i}_{\Bz}\;\epsilon^{i}_{\Bz}.
\end{equation}
Here, $\epsilon^{i}_{\B^{+(0)}}$ denotes the total reconstruction and selection efficiency for the $\Bp$ ($\Bz$) decays in category $i$. These efficiencies are determined using simulated samples and corrected to account for known differences in track reconstruction, hardware trigger efficiency, and PID response between data and simulation. 

The efficiency of the three-body decay \bpdstdspi depends on the phase space variables, $\mathbf{x}$. Since the simulated signal decay density is generated uniformly over $\mathbf{x}$, the results from the amplitude analysis are used to account for this dependency by correcting the \Bp efficiencies with the factor
\begin{equation}
	\epsilon_{\Bp} = \frac{\int\epsilon\left(\mathbf{x}\right)\; \vert\mathcal{A(\mathbf{x})}\vert^2\; {\rm d}\mathbf{x}}{\int |\mathcal{A(\mathbf{x})}|^2\;{\rm d}\mathbf{x}},
\end{equation}
where $\mathcal{A}(\mathbf{x})$ is the fitted decay amplitude,  $\epsilon(\mathbf{x})$ is the four-dimensional efficiency map parametrised using ANN, and integration is performed over the full phase space of the decay.

The analysis also measures the ratio of \bpdstdsstpi and \bpdstdspi branching fractions, denoted as $\mathcal{R}^{*}$. This is achieved by linking the $N^{i}_{\bpdstdsstpi}$ and $N^{i}_{\bpdstdspi}$ yields in the simultaneous fit as
\begin{equation}
	N^{i}_{\bpdstdsstpi} = \mathcal{R}^{*}\;N^{i}_{\bpdstdspi}\;\xi^i,
\end{equation}
where $\xi^i=\epsilon^i_{\bpdstdspi}/\epsilon^i_{\bpdstdsstpi}$ denotes the ratio of the total reconstruction and selection efficiencies of the \bpdstdspi and \bpdstdsstpi modes, determined from simulation. 

The ratios of branching fractions measured from the fit are determined to be \mbox{$\mathcal{R} = 0.173 \pm 0.006$}, and 
\mbox{$\mathcal{R}^{*} = 1.32 \pm 0.07$}, where the uncertainties are statistical only. The consistency between the Run 1, Run 2 and TOS, NotTOS datasets is tested by performing separate fits to the data in the four categories, showing good agreement within the statistical uncertainties on the relative branching fractions.

A number of systematic uncertainties affecting the ratios of branching fractions are considered. Table~\ref{tab:systbranching} presents the systematic uncertainties evaluated in per cent relative to the values of the ratios of branching fractions. 

The yields obtained from the fit depend on the model chosen to describe the signal and background invariant-mass distributions. To assess these effects, alternative models are employed, and the difference between the alternative and the baseline fits is taken as the systematic uncertainty. The fully reconstructed signal model uncertainty is evaluated using a DCB shape instead of the baseline sum of the Gaussian and DCB distributions. An additional uncertainty is associated with the values of the DCB tail parameters and the fraction between the Gaussian and DCB distributions. This contribution is estimated by propagating the corresponding uncertainties on the values obtained from the simulation and is found to be negligible. 

The uncertainty due to the combinatorial background model is obtained by replacing the baseline exponential distribution with a linear function. For the partially reconstructed $\Bp\to \Dstarm\D^{*+}_s\pip$ contribution, a bifurcated Gaussian shape is considered while the uncertainty due to the partially reconstructed shape in the $\Bz\to \Dstarm\Dsp$ sample is considered to be negligible. 

The uncertainty due to the estimated fractions of non-\Dsp contamination in the fully reconstructed signal peaks is obtained by varying them within their uncertainties while fitting the data, and taking the RMS deviation as the systematic uncertainty.

The uncertainties in the efficiency ratios account for simulation sample size as well as uncertainties related to PID response, trigger, and nonuniformity corrections. The uncertainty due to limited simulation sample size is accounted for by varying the efficiencies obtained from the simulated samples in the data fit and taking the RMS of the fitted $\mathcal{R}^{(*)}$ values as the uncertainty. The PID correction uncertainty is assessed by extracting efficiencies after transforming the PID responses in simulation with a modified kernel width used for density estimation. The trigger efficiency correction uncertainty is evaluated by computing efficiencies from simulation samples without any trigger efficiency correction. Systematic uncertainties are determined by comparing these results to the baseline.

Non-uniformity correction factor uncertainties are driven by the limited size of simulation samples and uncertainties in the four-dimensional efficiency profile parametrisation. These uncertainties are propagated to the uncertainty in the calculation of the $\mathcal{R}^{(*)}$ values. For the \bpdstdsstpi mode, where the correction is not applied, the uncertainty is conservatively assigned to be the maximum absolute value of the correction for the \bpdstdspi mode (4.8\%).

\begin{table}[h]
\caption{Relative systematic uncertainties on the $\mathcal{R}^{(*)}$ measurements. }
\label{tab:systbranching}
\centering
\begin{tabular}{l|cc}
Source of systematic uncertainty & $\delta\mathcal{R}$ [\%] & $\delta\mathcal{R}^{*}$ [\%] \\
\midrule
Fully reconstructed signal model & 3.5 & 8.4 \\
Combinatorial background model & 3.7 & 2.8 \\
Partially reconstructed signal model & 1.0 & 2.9 \\
Non-$D_s$ fraction estimation & 0.1 & 0.2 \\
Efficiency, PID correction & 0.6 & 1.2 \\
Efficiency, sample size & 0.6 & 0.5 \\
Efficiency, trigger correction & 1.2 & 0.3 \\
Efficiency non-uniformity correction & 2.3 & 4.8 \\
\end{tabular}
\end{table}

The total systematic uncertainty is taken as the sum in quadrature of each of the contributions described above. Including them, the measured values for the ratios of branching fractions are 
\begin{align*}
\mathcal{R} & = 0.173 \pm 0.006\pm 0.010, \\ 
\mathcal{R}^{*} & = 1.32 \pm 0.07\pm 0.14,
\end{align*}
where the first uncertainty is statistical, and the second is total systematic. 

\section{Conclusion}
\label{sec:Conclusion}

The decays of the $\Bp$ meson to the \dstdspi final state are studied in this analysis using proton-proton collision data obtained with the LHCb detector. The data are collected at centre-of-mass energies of 7, 8, and 13\tev, corresponding to a total luminosity of 9\invfb. The branching ratio of the \bpdstdspi decay with respect to that of the \bdstds decay is 
\[
  \mathcal{R} = \frac{\BR(\bpdstdspi)}{\BR(\bdstds)} = 0.173\pm 0.006\pm 0.010, 
\]
where the first uncertainty is statistical and the second is systematic. 
The partially reconstructed decay $\bpdstdsstpi$ (with $\D_{s}^{\ast +}\to \Dsp\gamma$ or $\Ds\piz$, where $\gamma$ or $\piz$ are not reconstructed) 
is also observed, and the ratio of branching fractions is measured to be
\[
  \mathcal{R}^{*} = \frac{\BR(\bpdstdsstpi)}{\BR(\bpdstdspi)} = 1.32\pm 0.07\pm 0.14. 
\]
The amplitude analysis of the \bpdstdspi decays has been performed for the first time. It was found that the amplitude is dominated by the following resonances in the $\Dstarm\pip$ channel: $D_1(2420)$, $D_1(2430)$, $D_2^*(2460)$, $D_0(2550)$, $D_1^*(2600)$, $D_2(2740)$. The fit fractions and phases for the components of the amplitude are presented in Table~\ref{tab:ff_results}. No strong evidence of exotic contributions 
in the $\Dstarm\Dsp$ or $\Dsp\pip$ channels is observed. The fit fraction of the scalar state $\tcsbar$ in the $\Dsp\pip$ channel observed in the $\Bp\to \Dm\Dsp\pip$ analysis~\cite{LHCb-PAPER-2022-026, LHCb-PAPER-2022-027} is found to be less than 2.3\% at 90\% CL.

\section*{Acknowledgements}
%
%
\noindent We express our gratitude to our colleagues in the CERN
accelerator departments for the excellent performance of the LHC. We
thank the technical and administrative staff at the LHCb
institutes.
We acknowledge support from CERN and from the national agencies:
CAPES, CNPq, FAPERJ and FINEP (Brazil); 
MOST and NSFC (China); 
CNRS/IN2P3 (France); 
BMBF, DFG and MPG (Germany); 
INFN (Italy); 
NWO (Netherlands); 
MNiSW and NCN (Poland); 
MCID/IFA (Romania); 
MICIU and AEI (Spain); 
SNSF and SER (Switzerland); 
NASU (Ukraine); 
STFC (United Kingdom); 
DOE NP and NSF (USA).
We acknowledge the computing resources that are provided by CERN, IN2P3
(France), KIT and DESY (Germany), INFN (Italy), SURF (Netherlands),
PIC (Spain), GridPP (United Kingdom), 
CSCS (Switzerland), IFIN-HH (Romania), CBPF (Brazil),
and Polish WLCG (Poland).
We are indebted to the communities behind the multiple open-source
software packages on which we depend.
Individual groups or members have received support from
ARC and ARDC (Australia);
Key Research Program of Frontier Sciences of CAS, CAS PIFI, CAS CCEPP, 
Fundamental Research Funds for the Central Universities, 
and Sci. \& Tech. Program of Guangzhou (China);
Minciencias (Colombia);
EPLANET, Marie Sk\l{}odowska-Curie Actions, ERC and NextGenerationEU (European Union);
A*MIDEX, ANR, IPhU and Labex P2IO, and R\'{e}gion Auvergne-Rh\^{o}ne-Alpes (France);
AvH Foundation (Germany);
ICSC (Italy); 
Severo Ochoa and Mar\'ia de Maeztu Units of Excellence, GVA, XuntaGal, GENCAT, InTalent-Inditex and Prog.~Atracci\'on Talento, CM (Spain);
SRC (Sweden);
the Leverhulme Trust, the Royal Society
 and UKRI (United Kingdom).

\newpage
\section*{Appendices}
\appendix
\section{Results of the baseline and alternative fits}
\label{apdx:resultfits}

Table~\ref{tab:aman_fit_results} shows the fitted parameters for the baseline \bpdstdspi amplitude model and four alternative models used to assess the model uncertainty: the real and imaginary values of the $LS$ couplings $c_{L,S}$ and line shape parameters ($m_0$, $\Gamma_0$ for the Breit--Wigner line shapes and $\alpha$ for the nonresonant exponential line shapes). The fit fractions of each amplitude component and the NLL values for each model are also reported. ``NR'' stands for exponential nonresonant line shape, the other components are represented by Breit--Wigner functions (see~\ref{sec:Formalism}). The masses and width of the Breit--Wigner states are fixed to the values given in Table~\ref{tab:excitedDPDG} except for the case of the model with the floated $\D_1(2430)$ parameters. The values of the cubic spline amplitude $f_i$ in the six spline knots for the QMI $1^+$ amplitude are shown separately in Table~\ref{tab:spline_fit_results}. 

\begin{table}
\centering
\caption{Fit results for the baseline and selected alternative $\Bp\to\Dstarm\Dsp\pip$ amplitude models. }
\label{tab:aman_fit_results}
\scalebox{0.75}{
\begin{tabular}{ll|ccccc} 
          &            &           & Floated      &      & NR                & NR, \\ 
Component & Parameter  & Baseline  & $D_1(2430)$  & QMI  & $D^*\pi$ $1^{-}$  & \tcsbar \\ \midrule
  & $-\ln\mathcal{L}$ & $-1670.16$ & $-1671.12$ & $-1674.18$ & $-1683.01$ & $-1694.94$ \\ 
\midrule 
  $D_1(2420)$ & ${\rm Re}(c_{0,1})$& $-0.088 \pm 0.033$ & $-0.08 \pm 0.05$ & $-0.06 \pm 0.04$ & $-0.085 \pm 0.034$ & $-0.101 \pm 0.035$ \\ 
 & ${\rm Im}(c_{0,1})$& $-0.22 \pm 0.06$ & $-0.26 \pm 0.06$ & $-0.26 \pm 0.05$ & $-0.22 \pm 0.05$ & $-0.22 \pm 0.06$ \\ 
 & ${\rm Re}(c_{2,1})$& $\phantom{-}1.0$ (fixed) & $\phantom{-}1.0$ (fixed) & $\phantom{-}1.0$ (fixed) & $\phantom{-}1.0$ (fixed) & $\phantom{-}1.0$ (fixed) \\ 
 & ${\rm Im}(c_{2,1})$& $\phantom{-}0.0$ (fixed) & $\phantom{-}0.0$ (fixed) & $\phantom{-}0.0$ (fixed) & $\phantom{-}0.0$ (fixed) & $\phantom{-}0.0$ (fixed) \\ 
 & $\mathcal{F}$(S-wave) & $\phantom{-}0.038 \pm 0.017$ & $\phantom{-}0.050 \pm 0.024$ & $\phantom{-}0.046 \pm 0.018$ & $\phantom{-}0.040 \pm 0.018$ & $\phantom{-}0.036 \pm 0.019$ \\ 
 & $\mathcal{F}$(D-wave) & $\phantom{-}0.71 \pm 0.04$ & $\phantom{-}0.71 \pm 0.05$ & $\phantom{-}0.651 \pm 0.022$ & $\phantom{-}0.71 \pm 0.04$ & $\phantom{-}0.66 \pm 0.04$ \\ 
\midrule 
  $D_1(2430)$ & ${\rm Re}(c_{0,1})$& $\phantom{-}1.64 \pm 0.16$ & $\phantom{-}1.44 \pm 0.25$  & - & $\phantom{-}1.60 \pm 0.16$ & $\phantom{-}1.90 \pm 0.20$ \\ 
 & ${\rm Im}(c_{0,1})$& $\phantom{-}0.23 \pm 0.17$ & $-0.06 \pm 0.24$  & - & $\phantom{-}0.32 \pm 0.17$ & $\phantom{-}0.05 \pm 0.24$ \\ 
 & ${\rm Re}(c_{2,1})$& $-0.33 \pm 0.26$ & $-0.35 \pm 0.23$  & - & $-0.36 \pm 0.28$ & $-0.11 \pm 0.26$ \\ 
 & ${\rm Im}(c_{2,1})$& $-0.05 \pm 0.13$ & $\phantom{-}0.03 \pm 0.15$  & - & $-0.02 \pm 0.13$ & $-0.09 \pm 0.14$ \\ 
 & $m_0$ [$\gev\,$]& $2.412$ (fixed) & $\phantom{-}2.378 \pm 0.025$  & - & $2.412$ (fixed) & $2.412$ (fixed) \\ 
 & $\Gamma_0$ [$\gev\,$]& $0.314$ (fixed) & $\phantom{-}0.24 \pm 0.06$  & - & $0.314$ (fixed) & $0.314$ (fixed) \\ 
 & $\mathcal{F}$(S-wave) & $\phantom{-}0.142 \pm 0.025$ & $\phantom{-}0.139 \pm 0.031$  & - & $\phantom{-}0.137 \pm 0.025$ & $\phantom{-}0.17 \pm 0.04$ \\ 
 & $\mathcal{F}$(D-wave) & $\phantom{-}0.005 \pm 0.009$ & $\phantom{-}0.007 \pm 0.012$  & - & $\phantom{-}0.006 \pm 0.008$ & $\phantom{-}0.001 \pm 0.004$ \\ 
\midrule 
  $1^+S$ QMI  & $\mathcal{F}$(S-wave)  & -  & - & $\phantom{-}0.122 \pm 0.024$  & -  & - \\ 
\midrule
  $D_2^*(2460)$ & ${\rm Re}(c_{2,1})$& $-0.64 \pm 0.04$ & $-0.64 \pm 0.05$ & $-0.67 \pm 0.04$ & $-0.62 \pm 0.05$ & $-0.66 \pm 0.05$ \\ 
 & ${\rm Im}(c_{2,1})$& $\phantom{-}0.00 \pm 0.07$ & $-0.00 \pm 0.07$ & $\phantom{-}0.03 \pm 0.07$ & $\phantom{-}0.01 \pm 0.07$ & $-0.03 \pm 0.07$ \\ 
 & $\mathcal{F}$(D-wave) & $\phantom{-}0.117 \pm 0.014$ & $\phantom{-}0.116 \pm 0.014$ & $\phantom{-}0.116 \pm 0.014$ & $\phantom{-}0.109 \pm 0.014$ & $\phantom{-}0.115 \pm 0.016$ \\ 
\midrule 
  $D_0(2550)$ & ${\rm Re}(c_{1,1})$& $-0.15 \pm 0.05$ & $-0.15 \pm 0.05$ & $-0.21 \pm 0.06$ & $-0.15 \pm 0.06$ & $-0.11 \pm 0.06$ \\ 
 & ${\rm Im}(c_{1,1})$& $-0.19 \pm 0.05$ & $-0.19 \pm 0.05$ & $-0.18 \pm 0.05$ & $-0.21 \pm 0.06$ & $-0.08 \pm 0.06$ \\ 
 & $\mathcal{F}$(P-wave) & $\phantom{-}0.023 \pm 0.008$ & $\phantom{-}0.023 \pm 0.008$ & $\phantom{-}0.026 \pm 0.008$ & $\phantom{-}0.027 \pm 0.009$ & $\phantom{-}0.007 \pm 0.005$ \\ 
\midrule 
  $D_1^*(2600)$ & ${\rm Re}(c_{1,1})$& $\phantom{-}0.53 \pm 0.07$ & $\phantom{-}0.54 \pm 0.07$ & $\phantom{-}0.54 \pm 0.07$ & $\phantom{-}0.34 \pm 0.09$ & $\phantom{-}0.50 \pm 0.09$ \\ 
 & ${\rm Im}(c_{1,1})$& $\phantom{-}0.17 \pm 0.09$ & $\phantom{-}0.17 \pm 0.09$ & $\phantom{-}0.19 \pm 0.10$ & $\phantom{-}0.25 \pm 0.12$ & $\phantom{-}0.41 \pm 0.11$ \\ 
 & $\mathcal{F}$(P-wave) & $\phantom{-}0.048 \pm 0.010$ & $\phantom{-}0.049 \pm 0.010$ & $\phantom{-}0.047 \pm 0.009$ & $\phantom{-}0.028 \pm 0.013$ & $\phantom{-}0.059 \pm 0.015$ \\ 
\midrule 
  $D_2(2740)$ & ${\rm Re}(c_{1,1})$& $\phantom{-}0.15 \pm 0.07$ & $\phantom{-}0.17 \pm 0.06$ & $\phantom{-}0.19 \pm 0.07$ & $\phantom{-}0.16 \pm 0.07$ & $\phantom{-}0.11 \pm 0.07$ \\ 
 & ${\rm Im}(c_{1,1})$& $-0.00 \pm 0.09$ & $\phantom{-}0.00 \pm 0.09$ & $\phantom{-}0.03 \pm 0.08$ & $\phantom{-}0.00 \pm 0.09$ & $-0.08 \pm 0.11$ \\ 
 & ${\rm Re}(c_{3,1})$& $\phantom{-}0.39 \pm 0.07$ & $\phantom{-}0.40 \pm 0.07$ & $\phantom{-}0.40 \pm 0.08$ & $\phantom{-}0.39 \pm 0.08$ & $\phantom{-}0.42 \pm 0.07$ \\ 
 & ${\rm Im}(c_{3,1})$& $-0.04 \pm 0.11$ & $-0.03 \pm 0.11$ & $-0.05 \pm 0.10$ & $-0.00 \pm 0.12$ & $-0.14 \pm 0.11$ \\ 
 & $\mathcal{F}$(P-wave) & $\phantom{-}0.004 \pm 0.004$ & $\phantom{-}0.004 \pm 0.004$ & $\phantom{-}0.005 \pm 0.005$ & $\phantom{-}0.004 \pm 0.004$ & $\phantom{-}0.003 \pm 0.004$ \\ 
 & $\mathcal{F}$(F-wave) & $\phantom{-}0.023 \pm 0.007$ & $\phantom{-}0.025 \pm 0.008$ & $\phantom{-}0.023 \pm 0.007$ & $\phantom{-}0.022 \pm 0.008$ & $\phantom{-}0.027 \pm 0.008$ \\ 
\midrule 
  NR $1^-$ & ${\rm Re}(c_{1,1})$ & -  & -  & - & $\phantom{-}0.053 \pm 0.018$  & - \\ 
 & ${\rm Im}(c_{1,1})$ & -  & -  & - & $\phantom{-}0.034 \pm 0.020$  & - \\ 
 & $\alpha$ [$\gev^{-2}\,$] & -  & -  & - & $\phantom{-}0.99 \pm 0.30$  & - \\ 
 & $\mathcal{F}$(P-wave)  & -  & -  & - & $\phantom{-}0.025 \pm 0.009$  & - \\ 
\midrule 
  NR $\Dsp\pip$ & ${\rm Re}(c_{0,0})$ & -  & -  & -  & - & $-0.006 \pm 0.008$ \\ 
 & ${\rm Im}(c_{0,0})$ & -  & -  & -  & - & $\phantom{-}0.021 \pm 0.014$ \\ 
 & ${\rm Re}(c_{1,1})$ & -  & -  & -  & - & $\phantom{-}0.063 \pm 0.035$ \\ 
 & ${\rm Im}(c_{1,1})$ & -  & -  & -  & - & $\phantom{-}0.007 \pm 0.013$ \\ 
 & ${\rm Re}(c_{2,2})$ & -  & -  & -  & - & $-0.062 \pm 0.033$ \\ 
 & ${\rm Im}(c_{2,2})$ & -  & -  & -  & - & $\phantom{-}0.034 \pm 0.026$ \\ 
 & $\alpha$  [$\gev^{-2}\,$] & -  & -  & -  & - & $-0.32 \pm 0.14$ \\ 
 & $\mathcal{F}$(S-wave)  & -  & -  & -  & - & $\phantom{-}0.028 \pm 0.014$ \\ 
 & $\mathcal{F}$(P-wave)  & -  & -  & -  & - & $\phantom{-}0.048 \pm 0.014$ \\ 
 & $\mathcal{F}$(D-wave)  & -  & -  & -  & - & $\phantom{-}0.019 \pm 0.010$ \\ 
\midrule 
  \tcsbar & ${\rm Re}(c_{1,1})$ & -  & -  & -  & - & $\phantom{-}0.17 \pm 0.05$ \\ 
 & ${\rm Im}(c_{1,1})$ & -  & -  & -  & - & $\phantom{-}0.04 \pm 0.07$ \\ 
 & $\mathcal{F}$(P-wave)  & -  & -  & -  & - & $\phantom{-}0.012 \pm 0.008$ \\ 
\end{tabular} 
}
\end{table}

\begin{table}
\centering
\caption{Fitted complex values $f_i$ of the $1^+$ S-wave amplitudes at the spline knots of the QMI model.}
\label{tab:spline_fit_results}
\begin{tabular}{l|cc} 
Parameter  & ${\rm Re}(f_i)$  & ${\rm Im}(f_i)$  \\ \midrule
 $f_0$ & $\phantom{-}0.2 \pm 2.8$   & $\phantom{-}0.7 \pm 2.4$ \\ 
 $f_1$ & $\phantom{-}0.9 \pm 1.2$   & $\phantom{-}1.7 \pm 0.9$ \\ 
 $f_2$ & $\phantom{-}0.7 \pm 0.6$   & $\phantom{-}2.5 \pm 0.4$ \\ 
 $f_3$ &          $-0.72 \pm 0.19$  & $\phantom{-}1.12 \pm 0.26$ \\ 
 $f_4$ & $\phantom{-}0.02 \pm 0.17$ & $\phantom{-}0.31 \pm 0.22$ \\ 
 $f_5$ & $-1.4 \pm 0.5$             & $-0.4 \pm 0.9$ \\ 
\end{tabular} 
\end{table}

\section{Systematic uncertainties for the baseline fit}
\label{apdx:unc}

Tables~\ref{tab:unc1} and \ref{tab:unc2} present the systematic and model uncertainties on each of the fitted parameter of the baseline model, together with the uncertainties on the fit fractions and relative phases of the fit components. The details on how these uncertainties are obtained are given in Section~\ref{sec:uncertainties}. 

\begin{table}[b]
\centering
\caption{Absolute systematic uncertainties on the fit parameters and fit fractions for the baseline model}
\label{tab:unc1}
\scalebox{0.74}{
\begin{tabular}{l|rrrrrrrrrrrr}
{} &  \begin{sideways}Accept., sample size\end{sideways} &  \begin{sideways}Accept., ANN param.\end{sideways} &  \begin{sideways}Efficiency, sample size\end{sideways} &  \begin{sideways}Efficiency, ANN param.\end{sideways} &  \begin{sideways}Efficiency, PID corr.\end{sideways} &  \begin{sideways}Efficiency, L0 corr.\end{sideways} &  \begin{sideways}Comb. bkg., sample size\end{sideways} &  \begin{sideways}Comb. bkg., ANN param.\end{sideways} &  \begin{sideways}Comb. bkg. fraction\end{sideways} &  \begin{sideways}Non-$D_s$ bkg., sample size\end{sideways} &  \begin{sideways}Non-$D_s$ bkg., ANN param.\end{sideways} &  \begin{sideways}Non-$D_s$ bkg. fraction\end{sideways} \\
\midrule
$D_1(2420)$ ${\rm Re}(c_{0,1})$       &                                              0.002 &                                              0.004 &                                              0.008 &                                              0.008 &                                              0.001 &                                              0.002 &                                              0.003 &                                              0.005 &                                              0.002 &                                              0.001 &                                              0.005 &                                              0.003 \\
$D_1(2420)$ ${\rm Im}(c_{0,1})$       &                                              0.003 &                                              0.004 &                                              0.006 &                                              0.007 &                                              0.005 &                                              0.002 &                                              0.005 &                                              0.011 &                                              0.003 &                                              0.002 &                                              0.010 &                                              0.002 \\
$D_1(2430)$ ${\rm Re}(c_{0,1})$       &                                              0.008 &                                              0.005 &                                              0.017 &                                              0.019 &                                              0.028 &                                              0.009 &                                              0.030 &                                              0.050 &                                              0.017 &                                              0.011 &                                              0.022 &                                              0.021 \\
$D_1(2430)$ ${\rm Im}(c_{0,1})$       &                                              0.005 &                                              0.003 &                                              0.012 &                                              0.013 &                                              0.005 &                                              0.007 &                                              0.014 &                                              0.069 &                                              0.010 &                                              0.008 &                                              0.002 &                                              0.008 \\
$D_1(2430)$ ${\rm Re}(c_{2,1})$       &                                              0.009 &                                              0.012 &                                              0.015 &                                              0.010 &                                              0.023 &                                              0.000 &                                              0.031 &                                              0.032 &                                              0.007 &                                              0.023 &                                              0.038 &                                              0.012 \\
$D_1(2430)$ ${\rm Im}(c_{2,1})$       &                                              0.005 &                                              0.004 &                                              0.009 &                                              0.008 &                                              0.003 &                                              0.001 &                                              0.014 &                                              0.014 &                                              0.001 &                                              0.008 &                                              0.019 &                                              0.001 \\
$D_2^*(2460)$ ${\rm Re}(c_{2,1})$     &                                              0.003 &                                              0.004 &                                              0.004 &                                              0.004 &                                              0.005 &                                              0.006 &                                              0.003 &                                              0.002 &                                              0.001 &                                              0.002 &                                              0.006 &                                              0.001 \\
$D_2^*(2460)$ ${\rm Im}(c_{2,1})$     &                                              0.003 &                                              0.006 &                                              0.009 &                                              0.007 &                                              0.003 &                                              0.001 &                                              0.005 &                                              0.006 &                                              0.001 &                                              0.002 &                                              0.005 &                                              0.002 \\
$D_0(2550)$ ${\rm Re}(c_{1,1})$       &                                              0.002 &                                              0.002 &                                              0.005 &                                              0.007 &                                              0.006 &                                              0.006 &                                              0.006 &                                              0.011 &                                              0.001 &                                              0.005 &                                              0.003 &                                              0.001 \\
$D_0(2550)$ ${\rm Im}(c_{1,1})$       &                                              0.004 &                                              0.003 &                                              0.008 &                                              0.010 &                                              0.004 &                                              0.001 &                                              0.006 &                                              0.014 &                                              0.001 &                                              0.005 &                                              0.005 &                                              0.001 \\
$D_1^*(2600)$ ${\rm Re}(c_{1,1})$     &                                              0.003 &                                              0.005 &                                              0.005 &                                              0.003 &                                              0.003 &                                              0.008 &                                              0.006 &                                              0.009 &                                              0.002 &                                              0.006 &                                              0.004 &                                              0.007 \\
$D_1^*(2600)$ ${\rm Im}(c_{1,1})$     &                                              0.004 &                                              0.002 &                                              0.008 &                                              0.011 &                                              0.010 &                                              0.007 &                                              0.008 &                                              0.019 &                                              0.001 &                                              0.006 &                                              0.007 &                                              0.003 \\
$D_2(2740)$ ${\rm Re}(c_{1,1})$       &                                              0.001 &                                              0.001 &                                              0.005 &                                              0.004 &                                              0.003 &                                              0.001 &                                              0.005 &                                              0.003 &                                              0.001 &                                              0.004 &                                              0.003 &                                              0.002 \\
$D_2(2740)$ ${\rm Im}(c_{1,1})$       &                                              0.002 &                                              0.001 &                                              0.006 &                                              0.005 &                                              0.004 &                                              0.006 &                                              0.011 &                                              0.016 &                                              0.002 &                                              0.003 &                                              0.009 &                                              0.003 \\
$D_2(2740)$ ${\rm Re}(c_{3,1})$       &                                              0.001 &                                              0.002 &                                              0.007 &                                              0.005 &                                              0.007 &                                              0.000 &                                              0.009 &                                              0.024 &                                              0.002 &                                              0.007 &                                              0.031 &                                              0.004 \\
$D_2(2740)$ ${\rm Im}(c_{3,1})$       &                                              0.004 &                                              0.003 &                                              0.009 &                                              0.006 &                                              0.007 &                                              0.005 &                                              0.010 &                                              0.020 &                                              0.001 &                                              0.011 &                                              0.008 &                                              0.002 \\
\midrule
$\mathcal{F}$($D_1(2420)$ S-wave)          &                                              0.001 &                                              0.001 &                                              0.002 &                                              0.002 &                                              0.001 &                                              0.000 &                                              0.001 &                                              0.003 &                                              0.001 &                                              0.001 &                                              0.003 &                                              0.001 \\
$\mathcal{F}$($D_1(2420)$ D-wave)          &                                              0.002 &                                              0.005 &                                              0.005 &                                              0.003 &                                              0.003 &                                              0.004 &                                              0.004 &                                              0.004 &                                              0.001 &                                              0.004 &                                              0.007 &                                              0.005 \\
$\mathcal{F}$($D_1(2430)$ S-wave)          &                                              0.001 &                                              0.001 &                                              0.003 &                                              0.003 &                                              0.004 &                                              0.001 &                                              0.005 &                                              0.007 &                                              0.002 &                                              0.001 &                                              0.002 &                                              0.002 \\
$\mathcal{F}$($D_1(2430)$ D-wave)          &                                              0.000 &                                              0.000 &                                              0.000 &                                              0.000 &                                              0.001 &                                              0.000 &                                              0.001 &                                              0.001 &                                              0.000 &                                              0.001 &                                              0.001 &                                              0.000 \\
$\mathcal{F}$($D_2(2460)$ D-wave)                 &                                              0.001 &                                              0.001 &                                              0.001 &                                              0.002 &                                              0.001 &                                              0.002 &                                              0.001 &                                              0.000 &                                              0.000 &                                              0.001 &                                              0.001 &                                              0.000 \\
$\mathcal{F}$($D_0(2550)$ P-wave)                 &                                              0.001 &                                              0.000 &                                              0.001 &                                              0.001 &                                              0.001 &                                              0.001 &                                              0.001 &                                              0.001 &                                              0.000 &                                              0.000 &                                              0.000 &                                              0.000 \\
$\mathcal{F}$($D_1(2600)$ P-wave)                 &                                              0.000 &                                              0.001 &                                              0.001 &                                              0.001 &                                              0.000 &                                              0.001 &                                              0.001 &                                              0.003 &                                              0.000 &                                              0.001 &                                              0.001 &                                              0.001 \\
$\mathcal{F}$($D_2(2740)$ P-wave)          &                                              0.000 &                                              0.000 &                                              0.000 &                                              0.000 &                                              0.000 &                                              0.000 &                                              0.000 &                                              0.000 &                                              0.000 &                                              0.000 &                                              0.000 &                                              0.000 \\
$\mathcal{F}$($D_2(2740)$ F-wave)          &                                              0.000 &                                              0.000 &                                              0.001 &                                              0.000 &                                              0.001 &                                              0.000 &                                              0.001 &                                              0.003 &                                              0.000 &                                              0.001 &                                              0.004 &                                              0.000 \\
\midrule
$D_{1}(2420)$ S-wave phase     &                                              0.010 &                                              0.024 &                                              0.029 &                                              0.031 &                                              0.005 &                                              0.012 &                                              0.017 &                                              0.036 &                                              0.012 &                                              0.004 &                                              0.004 &                                              0.012 \\
$D_{1}(2430)$ S-wave phase     &                                              0.003 &                                              0.002 &                                              0.007 &                                              0.007 &                                              0.005 &                                              0.005 &                                              0.009 &                                              0.046 &                                              0.007 &                                              0.005 &                                              0.001 &                                              0.006 \\
$D_{1}(2430)$ D-wave phase     &                                              0.017 &                                              0.015 &                                              0.027 &                                              0.027 &                                              0.016 &                                              0.004 &                                              0.045 &                                              0.042 &                                              0.003 &                                              0.028 &                                              0.076 &                                              0.004 \\
$D^{*}_{2}(2460)$ D-wave phase &                                              0.005 &                                              0.009 &                                              0.014 &                                              0.011 &                                              0.005 &                                              0.001 &                                              0.008 &                                              0.009 &                                              0.002 &                                              0.003 &                                              0.007 &                                              0.003 \\
$D_{0}(2550)$ P-wave phase     &                                              0.015 &                                              0.010 &                                              0.028 &                                              0.045 &                                              0.007 &                                              0.018 &                                              0.034 &                                              0.072 &                                              0.005 &                                              0.028 &                                              0.023 &                                              0.006 \\
$D^{*}_{1}(2600)$ P-wave phase &                                              0.006 &                                              0.005 &                                              0.013 &                                              0.017 &                                              0.018 &                                              0.016 &                                              0.014 &                                              0.026 &                                              0.001 &                                              0.012 &                                              0.014 &                                              0.002 \\
$D_{2}(2740)$ P-wave phase     &                                              0.014 &                                              0.008 &                                              0.038 &                                              0.030 &                                              0.030 &                                              0.041 &                                              0.070 &                                              0.107 &                                              0.014 &                                              0.019 &                                              0.061 &                                              0.019 \\
$D_{2}(2740)$ F-wave phase     &                                              0.010 &                                              0.007 &                                              0.022 &                                              0.017 &                                              0.016 &                                              0.012 &                                              0.024 &                                              0.054 &                                              0.003 &                                              0.032 &                                              0.030 &                                              0.005 \\
\end{tabular}
}
\end{table}
\clearpage

\begin{table}[b]
\centering
\caption{Absolute systematic uncertainties on the fit parameters and fit fractions for the baseline model (continued)}
\label{tab:unc2}
\scalebox{0.74}{
\begin{tabular}{l|rrrr|rrrr}
{} &  \begin{sideways}Normalis. sample size\end{sideways} &  \begin{sideways}Blatt-Weisskopf radius\end{sideways} &  \begin{sideways}Resonance parameters\end{sideways} &  \begin{sideways}Fit bias\end{sideways} &  \begin{sideways}Fitted value\end{sideways} &  \begin{sideways}Stat. uncert.\end{sideways} &  \begin{sideways}Total syst. uncert.\end{sideways} & \begin{sideways}Model uncert.\end{sideways} \\
\midrule
$D_1(2420)$ ${\rm Re}(c_{0,1})$       &                                              0.003 &                                              0.010 &                                              0.005 &                                   0.000 &                                      $-$0.088 &                                        0.033 &                                              0.018 &                           $-0.013$/$+0.029$ \\
$D_1(2420)$ ${\rm Im}(c_{0,1})$       &                                              0.003 &                                              0.012 &                                              0.004 &                                   0.000 &                                      $-$0.216 &                                        0.056 &                                              0.024 &                           $-0.046$/$+0.001$ \\
$D_1(2430)$ ${\rm Re}(c_{0,1})$       &                                              0.008 &                                              0.028 &                                              0.116 &                                   0.016 &                                       1.642 &                                        0.157 &                                              0.144 &                           $-0.202$/$+0.260$ \\
$D_1(2430)$ ${\rm Im}(c_{0,1})$       &                                              0.009 &                                              0.171 &                                              0.114 &                                   0.013 &                                       0.228 &                                        0.174 &                                              0.219 &                           $-0.285$/$+0.091$ \\
$D_1(2430)$ ${\rm Re}(c_{2,1})$       &                                              0.009 &                                              0.111 &                                              0.170 &                                   0.013 &                                      $-$0.325 &                                        0.261 &                                              0.217 &                           $-0.033$/$+0.213$ \\
$D_1(2430)$ ${\rm Im}(c_{2,1})$       &                                              0.007 &                                              0.257 &                                              0.298 &                                   0.003 &                                      $-$0.049 &                                        0.134 &                                              0.395 &                           $-0.046$/$+0.077$ \\
$D_2^*(2460)$ ${\rm Re}(c_{2,1})$     &                                              0.004 &                                              0.009 &                                              0.041 &                                   0.005 &                                      $-$0.641 &                                        0.045 &                                              0.045 &                           $-0.025$/$+0.020$ \\
$D_2^*(2460)$ ${\rm Im}(c_{2,1})$     &                                              0.003 &                                              0.053 &                                              0.071 &                                   0.007 &                                       0.002 &                                        0.071 &                                              0.090 &                           $-0.035$/$+0.027$ \\
$D_0(2550)$ ${\rm Re}(c_{1,1})$       &                                              0.003 &                                              0.018 &                                              0.033 &                                   0.004 &                                      $-$0.154 &                                        0.053 &                                              0.042 &                           $-0.053$/$+0.044$ \\
$D_0(2550)$ ${\rm Im}(c_{1,1})$       &                                              0.003 &                                              0.012 &                                              0.041 &                                   0.002 &                                      $-$0.194 &                                        0.052 &                                              0.048 &                           $-0.021$/$+0.110$ \\
$D_1^*(2600)$ ${\rm Re}(c_{1,1})$     &                                              0.003 &                                              0.013 &                                              0.066 &                                   0.008 &                                       0.528 &                                        0.066 &                                              0.070 &                           $-0.185$/$+0.013$ \\
$D_1^*(2600)$ ${\rm Im}(c_{1,1})$     &                                              0.003 &                                              0.048 &                                              0.065 &                                   0.020 &                                       0.175 &                                        0.089 &                                              0.088 &                           $-0.005$/$+0.234$ \\
$D_2(2740)$ ${\rm Re}(c_{1,1})$       &                                              0.001 &                                              0.007 &                                              0.015 &                                   0.007 &                                       0.153 &                                        0.065 &                                              0.021 &                           $-0.043$/$+0.033$ \\
$D_2(2740)$ ${\rm Im}(c_{1,1})$       &                                              0.001 &                                              0.004 &                                              0.018 &                                   0.007 &                                      $-$0.003 &                                        0.085 &                                              0.032 &                           $-0.073$/$+0.030$ \\
$D_2(2740)$ ${\rm Re}(c_{3,1})$       &                                              0.002 &                                              0.054 &                                              0.066 &                                   0.031 &                                       0.392 &                                        0.068 &                                              0.100 &                           $-0.006$/$+0.029$ \\
$D_2(2740)$ ${\rm Im}(c_{3,1})$       &                                              0.002 &                                              0.034 &                                              0.052 &                                   0.004 &                                      $-$0.036 &                                        0.107 &                                              0.069 &                           $-0.103$/$+0.032$ \\
\midrule
$\mathcal{F}$($D_1(2420)$ S-wave)          &                                              0.001 &                                              0.004 &                                              0.001 &                                   0.002 &                                       0.038 &                                        0.017 &                                              0.008 &                           $-0.001$/$+0.013$ \\
$\mathcal{F}$($D_1(2420)$ D-wave)          &                                              0.002 &                                              0.024 &                                              0.013 &                                   0.006 &                                       0.710 &                                        0.044 &                                              0.032 &                           $-0.060$/$+0.000$ \\
$\mathcal{F}$($D_1(2430)$ S-wave)          &                                              0.001 &                                              0.006 &                                              0.014 &                                   0.001 &                                       0.142 &                                        0.025 &                                              0.019 &                           $-0.020$/$+0.031$ \\
$\mathcal{F}$($D_1(2430)$ D-wave)          &                                              0.000 &                                              0.005 &                                              0.012 &                                   0.003 &                                       0.005 &                                        0.009 &                                              0.013 &                           $-0.005$/$+0.002$ \\
$\mathcal{F}$($D_2(2460)$ D-wave)                 &                                              0.001 &                                              0.001 &                                              0.002 &                                   0.002 &                                       0.117 &                                        0.014 &                                              0.005 &                           $-0.007$/$+0.000$ \\
$\mathcal{F}$($D_0(2550)$ P-wave)                 &                                              0.000 &                                              0.000 &                                              0.001 &                                   0.001 &                                       0.023 &                                        0.008 &                                              0.003 &                           $-0.017$/$+0.003$ \\
$\mathcal{F}$($D_1(2600)$ P-wave)                 &                                              0.000 &                                              0.005 &                                              0.003 &                                   0.001 &                                       0.048 &                                        0.010 &                                              0.007 &                           $-0.020$/$+0.011$ \\
$\mathcal{F}$($D_2(2740)$ P-wave)          &                                              0.000 &                                              0.000 &                                              0.000 &                                   0.002 &                                       0.004 &                                        0.004 &                                              0.002 &                           $-0.001$/$+0.001$ \\
$\mathcal{F}$($D_2(2740)$ F-wave)          &                                              0.000 &                                              0.006 &                                              0.002 &                                   0.001 &                                       0.023 &                                        0.007 &                                              0.008 &                           $-0.001$/$+0.004$ \\
\midrule
$D_{1}(2420)$ S-wave phase     &                                              0.011 &                                              0.040 &                                              0.018 &                                   0.001 &                                      $-$1.959 &                                        0.160 &                                              0.082 &                           $-0.053$/$+0.166$ \\
$D_{1}(2430)$ S-wave phase     &                                              0.005 &                                              0.103 &                                              0.054 &                                   0.009 &                                       0.138 &                                        0.105 &                                              0.127 &                           $-0.178$/$+0.059$ \\
$D_{1}(2430)$ D-wave phase     &                                              0.020 &                                              0.682 &                                              0.472 &                                   0.002 &                                      $-$2.992 &                                        0.419 &                                              0.837 &                           $-0.230$/$+0.551$ \\
$D^{*}_{2}(2460)$ D-wave phase &                                              0.005 &                                              0.083 &                                              0.111 &                                   0.011 &                                       3.138 &                                        0.110 &                                              0.141 &                           $-0.041$/$+0.053$ \\
$D_{0}(2550)$ P-wave phase     &                                              0.012 &                                              0.085 &                                              0.179 &                                   0.009 &                                      $-$2.242 &                                        0.214 &                                              0.225 &                           $-0.252$/$+0.049$ \\
$D^{*}_{1}(2600)$ P-wave phase &                                              0.006 &                                              0.078 &                                              0.108 &                                   0.031 &                                       0.319 &                                        0.156 &                                              0.145 &                           $-0.014$/$+0.371$ \\
$D_{2}(2740)$ P-wave phase     &                                              0.009 &                                              0.026 &                                              0.118 &                                   0.046 &                                      $-$0.022 &                                        0.558 &                                              0.207 &                           $-0.587$/$+0.163$ \\
$D_{2}(2740)$ F-wave phase     &                                              0.006 &                                              0.114 &                                              0.146 &                                   0.018 &                                      $-$0.091 &                                        0.270 &                                              0.204 &                           $-0.228$/$+0.082$ \\
\end{tabular}
}
\end{table}
\clearpage


\addcontentsline{toc}{section}{References}
\bibliographystyle{LHCb}
\bibliography{main,standard,LHCb-PAPER,LHCb-CONF,LHCb-DP,LHCb-TDR}

\newpage
\centerline
{\large\bf LHCb collaboration}
\begin{flushleft}
\small
R.~Aaij$^{35}$\lhcborcid{0000-0003-0533-1952},
A.S.W.~Abdelmotteleb$^{54}$\lhcborcid{0000-0001-7905-0542},
C.~Abellan~Beteta$^{48}$,
F.~Abudin{\'e}n$^{54}$\lhcborcid{0000-0002-6737-3528},
T.~Ackernley$^{58}$\lhcborcid{0000-0002-5951-3498},
J. A. ~Adams$^{66}$\lhcborcid{0009-0003-9175-689X},
A. A. ~Adefisoye$^{66}$\lhcborcid{0000-0003-2448-1550},
B.~Adeva$^{44}$\lhcborcid{0000-0001-9756-3712},
M.~Adinolfi$^{52}$\lhcborcid{0000-0002-1326-1264},
P.~Adlarson$^{79}$\lhcborcid{0000-0001-6280-3851}
% ...

{\footnotesize \it

% ...
$^{35}$Nikhef National Institute for Subatomic Physics, Amsterdam, Netherlands\\
% ...
$^{44}$Instituto Galego de F{\'\i}sica de Altas Enerx{\'\i}as (IGFAE), Universidade de Santiago de Compostela, Santiago de Compostela, Spain\\
% ...
$^{48}$Physik-Institut, Universit{\"a}t Z{\"u}rich, Z{\"u}rich, Switzerland\\
% ...
$^{52}$H.H. Wills Physics Laboratory, University of Bristol, Bristol, United Kingdom\\
% ...
$^{54}$Department of Physics, University of Warwick, Coventry, United Kingdom\\
% ...
$^{57}$School of Physics and Astronomy, University of Glasgow, Glasgow, United Kingdom\\
$^{58}$Oliver Lodge Laboratory, University of Liverpool, Liverpool, United Kingdom\\
% ...
$^{66}$Syracuse University, Syracuse, NY, United States\\
% ...
$^{79}$Department of Physics and Astronomy, Uppsala University, Uppsala, Sweden, associated to $^{57}$\\
% ...
}
\end{flushleft}

\end{document}